\documentclass[graybox, envcountchap]{svmult}

\usepackage{mathptmx}        
\usepackage{amsmath}
\usepackage{amssymb}
\usepackage{color}
\usepackage{helvet}          
\usepackage{courier}         
\usepackage{dirtree}

\usepackage{makeidx}        
\usepackage{graphicx}        
\usepackage{subfig}

\usepackage{multicol}        
\usepackage[bottom]{footmisc}

\usepackage{hyperref}        
\hypersetup{colorlinks=true,urlcolor=blue}

\usepackage[misc]{ifsym}

\makeindex             

\begin{document}


\title{General relativistic magnetohydrodynamics simulations for binary neutron star mergers}
\author{Kenta Kiuchi}
\institute{Kenta Kiuchi (\Letter) \at Max-Planck Institute for Gravitational Physics, Am M\"{u}hlenberg 2, 10559, Potsdam, Germany\email{kenta.kiuchi@aei.mpg.de}
}
%
%
\maketitle

\abstract{Binary neutron star mergers used to be the most promising candidate for gravitational waves for ground-based gravitational wave detectors, such as advanced LIGO and advanced VIRGO. This was proved by the detection of gravitational waves from a binary neutron star merger in 2017. Numerical modeling is pivotal in predicting and interpreting binary neutron star mergers. 
This chapter reviews the progress of fully general relativistic magnetized binary neutron star merger simulations. 
From 2008 to 2024, about forty numerical relativity simulations of magnetized binary neutron star mergers were conducted with a different level of sophistication. 
This chapter aims to comprehensively view the magnetohydrodynamics effect in binary neutron star mergers by reviewing all the related works. 
}

\section{Introduction}
After the observation of GW170817 associated with the electromagnetic counterparts GRB 170817A and AT 2017gfo~\cite{Goldstein:2017mmi,LIGOScientific:2017ync,LIGOScientific:2017vwq,Mooley:2018qfh,Savchenko:2017ffs}, binary neutron star mergers became a leading player in the multimessenger era. Numerical relativity is a chosen way to construct a theoretical modeling of binary neutron star mergers. All fundamental interactions are equally essential in binary neutron star mergers. Thus, the numerical relativity codes should self-consistently implement all the effects of the fundamental interactions. The numerical relativity community started making an effort to build a physical model of binary neutron star mergers before the GW170817 event from two aspects. One is the magnetohydrodynamics (MHD) effect in binary neutron star mergers. It is motivated by the binary pulsars' observational fact that neutron stars generally have magnetic fields~\cite{Lorimer:2008se}. The community is trying to figure out what the MHD effect is in binary neutron star mergers. 
The other aspect is microphysics, i.e., the neutrino emission process (see, for example, Refs.~\cite{Foucart:2022bth,Shibata:2019wef} as a review for the progress of this aspect). Since 2015, it has become feasible to perform a simulation that combines the two aspects, i.e., the MHD and neutrino effect~\cite{Palenzuela:2015dqa}. 

General relativistic magnetohydrodynamics (GRMHD) in full general relativity was initiated in 2005~\cite{Duez:2005sf,Shibata:2005gp}. 
Subsequently, a couple of numerical relativity codes successfully implemented GRMHD~\cite{Cipolletta:2019geh,Dionysopoulou:2012zv,Etienne:2010ui,Etienne:2015cea,Giacomazzo:2007ti,Kiuchi:2012qv,Most:2019kfe,Mosta:2013gwu,Neilsen:2014hha,Palenzuela:2012my,Palenzuela:2018sly}, and started to explore binary neutron star mergers. The modeling of magnetized binary neutron star mergers is classified into three categories: (i) Three-dimensional binary neutron star merger simulations, particularly focusing on the merger dynamics~\cite{Aguilera-Miret:2020dhz,Aguilera-Miret:2021fre,Aguilera-Miret:2023qih,Anderson:2008zp,Chabanov:2022twz,Ciolfi:2017uak,Ciolfi:2019fie,Ciolfi:2020hgg,Combi:2022nhg,Combi:2023yav,Dionysopoulou:2015tda,Endrizzi:2016kkf,Giacomazzo:2009mp,Giacomazzo:2010bx,Giacomazzo:2013uua,Giacomazzo:2014qba,deHaas:2022ytm,Kawamura:2016nmk,Kiuchi:2014hja,Kiuchi:2015sga,Kiuchi:2017zzg,Kiuchi:2022ubj,Kiuchi:2022nin,Kiuchi:2023obe,Liu:2008xy,Most:2023sft,Most:2023sme,Mosta:2020hlh,Palenzuela:2015dqa,Palenzuela:2021gdo,Palenzuela:2022kqk,Rezzolla:2011da,Ruiz:2016rai,Ruiz:2017inq,Ruiz:2017due,Ruiz:2019ezy,Ruiz:2020via,Ruiz:2021qmm,Sun:2022vri}. (ii) Simulation for binary neutron star merger remnants, in which the remnant is manually constructed by an equilibrium configuration or mapping from the three-dimensional merger simulations~\cite{Christie:2019lim,Duez:2005cj,Duez:2006qe,Fernandez:2018kax,Kiuchi:2012qv,Lehner:2011aa,Shibata:2005mz,Shibata:2011fj,Shibata:2021bbj,Shibata:2021xmo,Siegel:2013nrw,Siegel:2017jug,Siegel:2017nub,Stephens:2006cn}. (iii) Force-free simulation of inspiraling binary neutron star mergers, whose aim is to explore the precursor signal before the merger~\cite{Most:2020ami,Most:2022ojl,Palenzuela:2012my,Palenzuela:2013hu,Palenzuela:2013kra}. 
Since GW170817, the importance of a self-consistent modeling of binary neutron star mergers keeps growing because it is mandatory to compare numerical relativity simulations to observational data for interpreting/predicting gravitational wave events. 
Therefore, the numerical relativity community is responsible for conducting simulations that are quantitatively accurate enough for such a comparison. 
However, it is still far on the way. For this reason, in this chapter, we mainly review the category (i). 

After initiation of the GRMHD simulation in binary neutron star mergers, this field is rapidly evolving, but at the same time the situation is ``chaotic". Namely, there is no clear consensus among different numerical relativity groups on the role of MHD instabilities, jet launching, and neutron-rich matter ejection. Therefore, in this chapter, we intend to review all the related works with criticism as much as possible and try to seek a way to deepen our understanding of the MHD effect in binary neutron star mergers. 

Readers may find a detailed discussion of the formulation, methodology, and implementation in the other chapters. Also, readers may be interested in the review of GRMHD simulations of binary neutron star mergers~\cite{Ciolfi:2020cpf}. 
Despite its pioneering, we will not review Newtonian magnetized binary simulations. The notation in this section follows a standard notation in this field. For example, $B^i$, $R$, $\rho$, $\Omega$, and $v^i$ denote the magnetic field, the radius, the rest-mass density, the angular velocity, and velocity, respectively. 

The structure of this chapter is as follows. In Sec.~\ref{sec:chapte16-1}, we review a couple of crucial ingredients for the MHD effect in the context of binary neutron star mergers. Section~\ref{sec:chapter16_ideal_MHD} is devoted to the review of the magnetized binary neutron star merger simulations conducted to date. In Sec.~\ref{sec:chapter16_summary}, we summarize the current status of our understanding of the MHD effect in binary neutron star mergers and discuss prospects. 

\section{Magnetohydrodynamics instabilities, neutrino emission, and large-scale dynamo}\label{sec:chapte16-1}
This section reviews several relevant ingredients in binary neutron star mergers. The first one is magnetohydrodynamics (MHD) instabilities. We begin by estimating the magnetic-field energy in the pre-merger stage with the observed magnetic-field strength in the binary pulsars~\cite{Lorimer:2008se}.
\begin{align}
E_{\rm mag} \approx 4 \times 10^{41}~{\rm erg}~B_{12}^2 R_{6}^3,
\end{align}
where $B_{12}=B/10^{12}{\rm G}$ and $R_{6}=R/10^6{\rm cm}$, respectively. The magnetic field strength in the observed binary pulsars is in the range of $\sim 10^7$--$10^{12}~{\rm G}$~\cite{Lorimer:2008se}. The typical kinetic energy just before the merger is 
\begin{align}
E_\mathrm{kin} \approx 2\times 10^{53}~{\rm erg}~M_{2.7} v^2_{10},
\end{align}
where $M_{2.7}=M/2.7M_\odot$ and $v_{10}=v/10^{10}{\rm cm~s^{-1}}$. Therefore, the magnetic field is dynamically unimportant in the pre-merger stage unless we consider an unrealistically ultra-strong magnetic field. Several MHD instabilities are proposed, which could amplify the magnetic field efficiently in a short time scale up to the saturation level, implying that the magnetic field could become dynamically important during and after the merger. 

In Ref.~\cite{Rasio:1999ku}, Rasio and Shapiro first proposed the Kelvin-Helmholtz instability could amplify the magnetic field when the two neutron stars collide with each other, and the shear layer is formed at the contact interface unless the binary neutron star promptly collapses to a black hole~\cite{Shibata:2019wef}. Because the growth rate of this instability is proportional to the wavenumber, the small-scale vortices could grow in a much shorter time scale than the relevant dynamical timescale. These vortices could curl the magnetic field lines, which results in an efficient small-scale magnetic field amplification. Therefore, to confirm this picture in a grid-based simulation, the spatial grid resolution is a crucial ingredient. 

The Kelvin-Helmholtz instability phase is not expected to continue during an entire post-merger stage because the shock waves generated by the colliding motion of the two neutron stars would dissipate the shear layer. Namely, the shear layer would survive only for a particular timescale. The timescale strongly depends on binary neutron star models such as an equation of state of nuclear matter, the mass of the neutron star, and neutron star spin in the pre-merger stage. For example, if we consider a ``soft" equation of state, the approaching velocity becomes faster, compared to a ``stiff" equation of state, when the two neutron stars collide because the neutron stars are compact~\cite{Hotokezaka:2012ze}. Consequently, the timescale for which the shear layer survives becomes shorter. 
It implies that the grid resolution requirement to resolve the Kelvin-Helmholtz instability becomes challenging in such a case because we expect the instability would amplify the magnetic field up to the saturation level within the lifetime of the shear layer in reality. Pre-merger magnetic-field topology should be explored in the Kelvin-Helmholtz instability. However, a careful assessment is necessary to explore the saturation due to the Kelvin-Helmholtz instability. 

After the Kelvin-Helmholtz instability phase, the merger remnant could be subject to the magneto-rotational instability (MRI)~\cite{Balbus:1991ay} because it differentially rotates in general~\cite{Shibata:2006nm}. 
The typical MRI wavelength of the axisymmetric fastest-growing mode is 
\begin{align}
\lambda_\text{MRI} = \frac{B_p}{\sqrt{4\pi \rho}} \frac{2\pi}{\Omega} \approx 8 \times 10^{3}~{\rm cm}~B_{p,15} \rho_{15}^{1/2} \Omega_{8000}^{-1}, \label{eq:chaper16_MRI}
\end{align}
where $B_{p,15}=B_p/10^{15}~{\rm G}$, $\rho_{15}=\rho/10^{15}~{\rm g~cm^{-3}}$, and $\Omega_{8000}=\Omega/8000~{\rm rad~s^{-1}}$. In reality, the Kelvin-Helmholtz instability saturation could initialize the poloidal magnetic-field strength $B_p$ in the MRI phase. In a grid-based numerical relativity simulation, the spatial grid resolution is again a key ingredient in properly resolving the MRI. MRI quality factor $Q_\mathrm{MRI}$ quantifies the ability of the employed grid resolution to resolve the MRI:
\begin{align}
Q_\mathrm{MRI} = \frac{\lambda_\mathrm{MRI}}{\Delta x}. \label{eq:chapter16_MRI_Qfac}
\end{align}
The critical MRI quality factor below which the MRI-driven turbulence cannot be sustained is $\approx 8$--$10$~\cite{Hawley:2011tq}. It should be noted that without estimating the MRI quality factor, it is hard to quantify how the MRI turbulence is sustained (see also the discussion about $Q_{\rm MRI}$ in the binary neutron star merger context below). 

One caveat on this estimate based on the ideal MRI assumption is that the neutrino viscosity and drag could significantly suppress the MRI as diffusive and damping processes in reality~\cite{Guilet:2016sqd}. The neutrino viscosity (drag) becomes relevant when the neutrino mean free path becomes shorter (longer) than $\lambda_{\rm MRI}$. The dispersion relation of the viscous- and damping-MRI are
\begin{align}
\left[\left(\tilde{\sigma}_{\rm vis}+\tilde{k}^2_{\rm vis}\tilde{\nu}_\nu\right)\tilde{\sigma}_{\rm vis}+\tilde{k}^2_{\rm vis}\right]^2+\tilde{\kappa}^2\left[\tilde{\sigma}^2_{\rm vis}+\tilde{k}^2_{\rm vis}\right]-4\tilde{k}^2_{\rm vis}=0, \label{eq:chaper16_dis1} 
\end{align}
and
\begin{align}
\left[\left(\tilde{\sigma}_{\rm drag}+\tilde{\Gamma}_\nu\right)\tilde{\sigma}_{\rm drag}+\tilde{k}^2_{\rm drag}\right]^2+\tilde{\kappa}^2\left[\tilde{\sigma}^2_{\rm drag}+\tilde{k}^2_{\rm drag}\right]-4\tilde{k}^2_{\rm drag}=0, \label{eq:chaper16_dis2}
\end{align}
respectively, where $\tilde{\sigma}_{\rm vis/drag}=\sigma_{\rm vis/drag}/\Omega$, $\tilde{k}_{\rm vis/drag}=k_{\rm vis/drag}v_A/\Omega$, $\tilde{\kappa}^2=\kappa^2/\Omega^2$, $\tilde{\nu}_\nu=\nu_\nu\Omega/v_A^2$, and $\tilde{\Gamma}_\nu=\Gamma_\nu/\Omega$. $\sigma_{\rm vis/drag}$ and $k_{\rm vis/drag}$ are the growth rate and wave number of MRI. $v_A=B^z_\mathrm{hyp}/\sqrt{4\pi\rho}$ is the Alfv\'{e}n wave speed with the hypothetical poloidal magnetic-field strength $B^z_\mathrm{hyp}$. $\kappa^2$ is the epicyclic frequency. $\nu_\nu$ and $\Gamma_\nu$ are the neutrino viscosity and drag damping rates, respectively. 
Reference~\cite{Guilet:2016sqd} calibrated them in a one-dimensional supernova simulation with an {\it ab initio} neutrino radiation transport:
\begin{align}
&\nu_\nu = 1.2\times 10^{10}~\mathrm{cm^2~s^{-1}}~\rho_{13}^{-2}T_{10}^2,\\
&\Gamma_\nu=6\times 10^3~\mathrm{s^{-1}} T_{10}^6,
\end{align}
where $\rho_{13}=\rho/10^{13}~{\rm g~cm^{-3}}$ and $T_{10}=T/10~{\rm MeV}$. The neutrino mean free path $l_\nu$ is also calibrated by
\begin{align}
l_\nu = 2\times 10^3~\rm{cm}~\rho_{13}^{-1}T_{10}^{-2}. 
\end{align}
Therefore, with microphysics, i.e., finite temperature effect and neutrino emission, it is possible to estimate the growth rate of the non-ideal MRI. It suggests the importance of implementing microphysics in magnetized binary neutron star merger simulations.

It should be noted that the MRI is relevant to sustain the magneto-turbulence inside the merger remnant because the resultant turbulent viscosity facilitates the angular momentum transport and heats the matter inside the merger remnant. In other words, the MRI's linear, exponential growing phase could be irrelevant in the binary merger context unless the Kelvin-Helmholtz instability is suppressed. 

The Shakura-Sunyaev parameter is the other important diagnostics to quantify the effective turbulent viscosity driven by MRI~\cite{Shakura:1973}:
\begin{align}
\alpha_{SS} \equiv -\left\langle\frac{b^{(r)}b_{(\phi)}}{4\pi P}\right\rangle,
\end{align}
where $P$ is the pressure, and $b_{(i)}$ is the tetrad component of the magnetic field measured in the fluid-rest frame. $\langle\cdot\rangle$ denotes the time ensemble. 

Besides the turbulent viscosity, MRI-driven turbulence would play another important role. Since the Kelvin-Helmholtz instability could generate a strong, but {\it small-scale}, magnetic field in a short timescale, a mechanism is necessary to convert such a small-scale field to a large one for the magnetic field to be dynamically important inside the merger remnant. The mean-field dynamo theory, in which each physical quantity $Q$ is decomposed into the mean component $\bar{Q}$ and fluctuating component $q$, i.e., $Q=\bar{Q}+q$, is a candidate for such a mechanism. We assume that an average of $Q$ in the azimuthal direction gives the mean field $\bar{Q}$ in the binary neutron star merger context because the merger remnant has an approximately axisymmetric structure. Thus, in the framework of the ideal MHD, we cast the induction equations as
\begin{align}
\partial_t \bar{\bf{B}} = \bf{\nabla}\times\left(\bar{\bf{V}}\times \bar{\bf{B}}+\bar{\bf{\mathcal E}}\right), \label{eq:chapter16_induction}
\end{align}
where $\bar{\mathcal E}=\overline{\bf{v}\times\bf{b}}$ is the electromotive force generated by the fluctuations. $\bf{B}=\bar{\bf{B}}+\bf{b}$ is the magnetic field and $\bf{V}=\bar{\bf{V}}+\bf{v}$ is the velocity field.

The simplest mean-field dynamo is $\alpha\Omega$ dynamo in the context of Solar physics, and we express the electromotive force as a function of the mean magnetic field:
\begin{align}
\bar{\mathcal {E}}_i = \alpha_{ij} \bar{B}_j + \beta_{ij} \left(\overline{\bf{\nabla}\times \bar{\bf{B}}}\right)_j, \label{eq:chapter16_alphaomega}
\end{align}
where $\alpha_{ij}$ and $\beta_{ij}$ are tensors that should not depend on the mean magnetic field~\cite{Brandenburg:2004jv}. The former tensor is called dynamo $\alpha$, and the second term is turbulent resitivity. If we assume the diagonal component of the first term on the right-hand side is dominant and a purely cylindrical differential rotation $\bar{\bf{V}}=R\Omega \bf{e}_\phi$, we can reduce Eq.~(\ref{eq:chapter16_induction}) into
\begin{align}
&\partial_t \bar{B}_R = - \partial_z \left[\left(\bar{\bf V}\bf{\times}\bar{\bf B}\right)_\phi + \bar{\mathcal E}_\phi\right]\approx-\partial_z\left(\alpha_{\phi\phi}\bar{B}_\phi\right), \label{eq:chapter16_alphaomega1} \\
&\partial_t \bar{B}_z = \partial_R \left[\left(\bar{\bf V}\bf{\times}\bar{\bf B}\right)_\phi + \bar{\mathcal E}_\phi\right] \approx \partial_R\left(\alpha_{\phi\phi}\bar{B}_\phi\right), \label{eq:chapter16_alphaomega2}\\
&\partial_t \bar{B}_\phi = R \bar{B}_A \partial_A \Omega, \label{eq:chapter16_alphaomega3}
\end{align}
where we work in the cylindrical coordinates, $\partial_\phi=0$, and $A=R,z$. 

These equations tell us that the mean poloidal magnetic field $\bar{B}_A$ is generated by the electromotive force described by the dynamo $\alpha$ and the mean toroidal field $\bar{B}_\phi$. It is called the $\alpha$ effect. The mean toroidal field $\bar{B}_\phi$ is generated by the differential rotation and the mean poloidal field $\bar{B}_A$. It is called the $\Omega$ effect or merely magnetic winding. Therefore, if it is realized inside the binary neutron star merger remnant, the dynamo cycle is closed, and the large-scale (mean-field) magnetic field is generated. In the binary neutron star merger context, the MRI could produce and sustain the turbulence, i.e., $\bf{v}$ and $\bf{b}$, which is the key to generating the electromotive force $\bar{\mathcal{E}}$. Since the turbulence is easily killed by the numerical diffusion in a simulation, the high-resolution simulation is the key to exploring the magnetic field evolution in the binary neutron star merger context. 

We should mention the relevance of the convergence study for the spatial grid resolution to explore the MHD effect in binary neutron star mergers. We begin by estimating the magnetic winding timescale originating from a pre-merger large-scale magnetic field. Suppose we consider a binary neutron star merger with the highly magnetized end of $10^{12}$~G in the observed binary pulsars. The winding timescale $t_A$ is 
\begin{align}
t_A \sim R/v_A \sim 100~{\rm s}~\bar{B}_{R,12}^{-1}\rho_{15}^{1/2} R_{6}, \label{eq:chapter16_winding}
\end{align}
where $\bar{B}_{R,12}=\bar{B}_R/10^{12}~{\rm G}$. Therefore, the magnetic winding originating from the pre-merger large-scale magnetic field should be irrelevant during the post-merger evolution within the timescale of $O(10)$~s. 

However, the existing general relativistic magnetized binary neutron star merger simulations employed the pre-merger magnetic field of $10^{15}$--$10^{17}~{\rm G}$ except for the simulations in Refs.~\cite{Endrizzi:2016kkf,Giacomazzo:2009mp,Giacomazzo:2010bx,Kawamura:2016nmk} and the simulations employing a sub-grid model~\cite{Aguilera-Miret:2020dhz,Aguilera-Miret:2021fre,Aguilera-Miret:2023qih,Giacomazzo:2014qba,Palenzuela:2021gdo,Palenzuela:2022kqk}. The primarily reason for using such an unrealistic magnetic field strength is to compensate the extensive computational cost to resolve the Kelvin-Helmholtz instability and the MRI (see Eq.~(\ref{eq:chaper16_MRI})). However, the trade-off is to shorten the winding timescale originating from the assumed pre-merger large-scale magnetic field. $t_A$ could become comparable to or shorter than the timescale relevant for the post-merger evolution (see Eq.~(\ref{eq:chapter16_winding})). Therefore, if we employ a {\it single} grid resolution with the pre-merger large-scale magnetic field of $10^{15}$--$10^{17}~{\rm G}$, we can not disentangle whether the magnetic winding originating from such a large-scale magnetic field determines the subsequent evolution of the merger remnant or not. Since the magnetic winding is relatively easy to resolve numerically, the convergence study is essential to disentangle the abovementioned point. Namely, suppose the magnetic winding originating from the pre-merger large-scale field plays a primary role in the subsequent merger remnant evolution. In that case, the result should not be significantly different in the different resolution simulations. If resolving the small-scale turbulence is essential, like the MRI-driven $\alpha\Omega$ dynamo mentioned above, the convergence study would give a significantly different result.

Therefore, in the following section, we review the previous works by explicitly mentioning the grid resolution, the pre-merger large-scale magnetic field, and the level of sophistication of the microphysics. 



\section{Magnetohydrodynamics simulations}\label{sec:chapter16_ideal_MHD}
In this section, we review the general relativistic magnetized binary neutron star merger simulations in the framework of the ideal and resistive MHD approximation. To clarify how our community has deepened our understanding of the MHD effect in binary neutron star mergers, we review all the related works chronologically. 

Before starting the review, we should mention the pre-merger magnetic field topology widely used in the numerical relativity community. The field topology is called a confined field configuration:
\begin{align}
A_\phi = A_b \max\left(P-P_{\rm cut},0\right)^{n_s},
\end{align}
where $A_\phi$ is the toroidal component of the vector potential, $P$ is the pressure, $P_{\rm cut}$ is the cut-off pressure parameterized by a fraction of the maximum pressure, and the concentration parameter $n_s$. $A_b$ is chosen so that the desired magnetic field strength is achieved. 
Together with the assumption of no poloidal vector potential, $A_r=A_\theta=0$, it gives a confined purely poloidal magnetic field. Since it has been revealed that the simulation result is not sensitive, at least qualitatively, to the choice of $P_{\rm cut}$ and $n_s$, the default setting for the pre-merger magnetic field topology is the confined configuration in this review without specifying $P_{\rm cut}$ and $n_s$. Otherwise, we explicitly explain the magnetic field topology. 

Merger remnant of binary neutron star merger is classified into three classes; a prompt collapse case, a short-lived case, and long-lived case~\cite{Shibata:2019wef}. For the prompt collapse case, the merger remnant collapses to a black hole immediately after the merger. For the short-lived case, the remnant massive neutron star would survive within $O(0.01)~{\rm s}$, whose timescale would be shorter than the MRI-driven turbulent viscous timescale, the magnetic winding timescale, and the neutrino cooling timescale.~\footnote{The neutrino cooling timescale is estimated by $t_{\nu,\rm cool}\sim E_{{\rm thr},53}/L_{\nu,53}\sim O(1)~{\rm s}$ where $E_{{\rm thr},53}=E_{\rm thr}/10^{53}~{\rm erg}$ and $L_{\nu,53}=L_\nu/10^{53}~{\rm erg/s}$ are the thermal energy and the neutrino luminosity, respectively.} For the long-lived case, the merger remnant massive neutron star survives for $O(1$-$10)~{\rm s}$. 

Anderson and his collaborators initiated a general relativistic magnetized symmetric binary neutron star merger simulation~\cite{Anderson:2008zp}. They employed the grid resolution of $\Delta x=460~{\rm m}$ and the pre-merger magnetic field of $9.6\times 10^{15}~{\rm G}$. The neutron star is modeled by the polytropic equation of state with $\Gamma=2$. They concluded the Kelvin-Helmholtz instability amplifies the magnetic field. They also reported the emergence of the Taylor instability with $m=1$ mode and the magnetic buoyancy instability. They quantified how the magnetic field modifies the post-merger gravitational waveforms. 
However, because of the lack of a convergence study, there was room to be improved in understanding whether the Kelvin-Hemoholtz instability really activates or not. 

Subsequently, Liu and his collaborators simulated magnetized symmetric and asymmetric binary neutron star mergers in full general relativity~\cite{Liu:2008xy}. They employed the grid resolution of $\Delta x \lesssim 700~{\rm m}$, and the pre-merger magnetic field of $10^{16}~{\rm G}$.~\footnote{Precisely speaking, their simulation result is scale-free because of the normalization with the polytropic constant $K$.} The neutron star is modeled by the $\Gamma$-law with $\Gamma=2$. 
Their conclusion is summarized as follows: (i) For the hypermassive neutron star formation case, the MHD effect can cause observable differences in the dynamics and gravitational waves, but they are not as significant as those reported in Ref.~\cite{Anderson:2008zp}. The difference might originate from the initial data of the binary neutron star. On the one hand, in Ref.~\cite{Liu:2008xy}, the quasi-equilibrium configuration was employed as the initial data. 
On the other hand, Ref.~\cite{Anderson:2008zp} set up the initial data by superposing boosted metrics of two spherical neutron stars. 
(ii) For the prompt black hole formation case, the MHD effect plays a minor role in the case of the symmetric binary. For the asymmetric binary, the MHD effect enhances the disk mass. 
However, since their employed grid resolution is not fine enough to capture the Kelvin-Helmholtz instability and MRI, as they explicitly mentioned, the role of these instabilities was not explored. Also, since their simulation time is shorter than the magnetic winding timescale $t_A$, the role of the magnetic winding effect needs to be more clarified. 

Giacomazzo and his collaborators explored how the pre-merger magnetic field is imprinted in the inspiral gravitational waveforms~\cite{Giacomazzo:2009mp}. They employed the grid resolution of $\Delta x\approx 350~{\rm m}$, and varied the pre-merger magnetic field from $0$ to $10^{17}~{\rm G}$. The neutron star is modeled by the $\Gamma$-law equation of state with $\Gamma=2$. 
Their conclusion is that the inspiral gravitational waves are essentially indistinguishable from those in the unmagnetized case unless the pre-merger magnetic field is as strong as $10^{17}~{\rm G}$. They also investigated the magnetic field amplification via the Kelvin-Helmholtz instability in the shear layer and concluded that the toroidal field is exponentially amplified until it becomes as strong as the poloidal field. 
In Ref.~\cite{Giacomazzo:2010bx}, they extended their work by employing the grid resolution of $\Delta x = 221~{\rm m}$ and varying the pre-merger magnetic field from $10^8$ to $10^{12}~{\rm G}$. The equation of state is the same as that in their previous work. They also, for the first time, performed the convergence test by changing the grid resolutions $\Delta x=177~{\rm m}$ and $354~{\rm m}$. They reported the growth rate of the magnetic field due to the Kelvin-Helmholtz instability is $\sim 0.5~{\rm ms^{-1}}$, and it does not significantly depend on the grid resolution. They argued the potential reason for this is the shortness of the lifetime of the shear layer. In their particular model, it is $\approx 1~{\rm ms}$. Therefore, the efficiency of the Kelvin-Helmholtz instability is still a riddle. Also, since the employed resolution cannot capture the MRI~(\ref{eq:chaper16_MRI}) and the lifetime of the remnant massive neutron star in their ``high-mass binaries" is shorter than the magnetic winding timescale~(\ref{eq:chapter16_winding}), the MHD effect in binary neutron star mergers needs to be more clarified.  

Rezzolla and his collaborators reported the emergence of a jet-like structure in Ref.~\cite{Rezzolla:2011da}. They employed the pre-merger magnetic field of $10^{12}~{\rm G}$. The grid resolution was likely $\Delta x=221~{\rm m}$. Their model collapsed to a black hole after $\approx 7$--$8~{\rm ms}$, and the subsequent massive torus formation occurred. They argued that the magnetic field exponentially grows inside the massive torus, and the ordered-magnetic field line is generated by a non-linear dynamo. Then, the jet-like structure is produced, which can power short gamma-ray bursts. 
Although detailed discussions and analyses about the MRI and the dynamo mechanism were absent, the paper has the merit of showing 
that the binary neutron star merger with a magnetic field may explain some aspects of the gamma-ray burst phenomenology. 

Giacomazzo and Perna reported a stable magnetar formation in a binary neutron star merger~\cite{Giacomazzo:2013uua}. They employed the pre-merger magnetic field of $10^{12}~{\rm G}$ and the single grid resolution with $\Delta x =225~{\rm m}$. The NS is modeled with the $\Gamma$-law equation with $\Gamma=2.75$. Since they chose a model in which the total mass is below the maximum mass of the TOV star, the resultant merger remnant neutron star is stable. 
They found the inefficient amplification via the Kelvin-Helmholtz instability, which is consistent with their previous finding~\cite{Giacomazzo:2010bx} and subsequent amplification via the magnetic winding. Although the magnetic field did not reach the magnetar strength of $\sim 10^{14}$--$10^{15}~{\rm G}$ at the end of the simulation, approximately $\approx 55~{\rm ms}$ after the merger, they claimed the stable magnetar formation is possible because it is expected the magnetic winding and unresolved MRI would amplify the magnetic field furthermore. 

\begin{figure}[b]
\includegraphics[scale=.75]{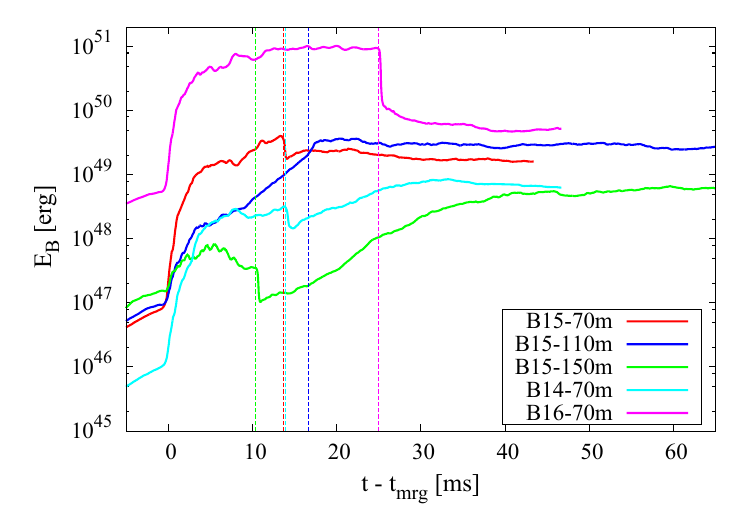}

\caption{Magnetic-field energy as a function of the post-merger time from Ref.~\cite{Kiuchi:2014hja}. 
$t-t_{\rm mrg}=0$ corresponds to the merger. B14, B15, and B16 denote the pre-merger magnetic field of $10^{14.5}~{\rm G}$, $10^{15}~{\rm G}$, and $10^{16}~{\rm G}$, respectively. The vertical dashed lines denote the black hole formation. The figure is taken from Ref.~\cite{Kiuchi:2014hja}.}
\label{fig:chapter16_Kiuchi2014} 
\end{figure}

Kiuchi and his collaborators reported magnetized binary neutron star merger simulations~\cite{Kiuchi:2014hja}. They employed the pre-merger magnetic field with $10^{14.5}~{\rm G}$, $10^{15}~{\rm G}$, and $10^{16}~{\rm G}$. They performed an in-depth convergence study by employing the grid resolution $\Delta x =70~{\rm m}$, $110~{\rm m}$, and $150~{\rm m}$. The neutron star is modeled with the H4 equation of state~\cite{Glendenning:1991es}. They employed the piecewise polytrope prescription for the cold part~\cite{Read:2008iy} and $\Gamma$-law equation for the thermal part. The merger remnant is short-lived. Contrary to the previous works~\cite{Giacomazzo:2010bx,Giacomazzo:2013uua,Rezzolla:2011da}, they found that an efficient magnetic-field amplification via the Kelvin-Helmholtz instability. Namely, the higher grid resolution simulation exhibits a larger growth rate of the magnetic field energy. It is consistent with the property of the Kelvin-Helmholtz instability, i.e., the shorter wavelength mode has the larger growth rate. However, even with their highest grid resolution, the saturation of the magnetic field is not achieved (see Fig.~\ref{fig:chapter16_Kiuchi2014}). Also, they, for the first time, estimated the MRI quality factor of the non-axisymmetric mode in the magnetized binary neutron star merger simulations. 
They confirmed that once the MRI quality factor exceeds the critical value of $10$, the magnetic field exponentially increases inside the merger remnant (see Fig.~3 in Ref.~\cite{Kiuchi:2014hja}). 
They also reported that even with their highest grid resolution, the MRI in a high-density region with $\rho \ge 10^{13}~{\rm g~cm^{-3}}$ cannot be captured because of the shortness of the MRI wavelength. They also reported the absence of the jet launching until the end of the simulation of $\approx 50~{\rm ms}$ after the black hole formation.

Giacomazzo and his collaborators reported new magnetized binary neutron star merger simulations by employing a sub-grid model~\cite{Giacomazzo:2014qba}. 
The pre-merger magnetic field is $\sim 2.5\times 10^{12}~{\rm G}$. The grid resolutions are $\approx 180~{\rm m}$, $220~{\rm m}$, $300~{\rm m}$, and $360~{\rm m}$. The neutron star is modeled with the $\Gamma$-law equation of state with $\Gamma=2$. They implemented a sub-grid model by adding a new term $\bf{E}_{\rm subgrid}$ in the induction equation:
\begin{align}
\partial_t \bf{A} = - \bf{E}_{\rm ideal} - \bf{E}_{\rm subgrid}, \label{eq:chapter16_subgrid}
\end{align}
where $\bf{A}$ is the vector potential and $\bf{E}_{\rm ideal}$ is the electric field in the ideal MHD. By assuming the closure relation between $\bf{E}_{\rm subgrid}$ and $\bf{A}$,
\begin{align}
{\bf E_{\rm subgrid}} = - S_{\rm subgrid} {\bf A}, \label{eq:chapter16_subgrid2}
\end{align}
where $S_{\rm subgrid}$ is a parameter, they reproduced an exponential amplification at the merger. 
To mimic the Kelvin-Helmholtz instability, i.e., the energy transfer from the turbulent kinetic energy in the shear layer to the magnetic field energy, and to suppress unphysical behavior near the stellar surface, they introduced four parameters in $S_{\rm subgrid}$. 
Two of them are proportional to the fraction of the fluid vorticity $\bf{\nabla} \times \bf{v}$, which are calibrated by high-resolution special relativistic MHD local box simulation for MHD turbulence~\cite{Zrake:2013mra}. 
They performed the convergence study and concluded that the saturation of the magnetic field is $\sim 10^{51}~{\rm erg}$. 
However, it is still an open question that the saturation is physical because it is controlled by $S_{\rm subgrid}$, which is calibrated only by special relativistic MHD. Furthermore, their sub-grid model violates energy conservation. 
Also, their closure relation needs to be justified. 

Dionysopoulou and her collaborators reported resistive MHD simulation of binary neutron star mergers for the first time~\cite{Dionysopoulou:2015tda}. 
They proposed a simplified model of the electrical conductivity $\sigma_{\rm con}$:~\footnote{Original notation for the electrical conductivity is $\sigma$, but to avoid confusion with the magnetization parameter $\sigma$, we use a different notation in this review.}
\begin{align}
\sigma_{\rm con}=\sigma_0 \max\left[1-\frac{2}{1+\exp\left[2D_{\rm tol}\left(D-D_{\rm rel}\right)/D_{\rm atm}\right]},0\right],
\end{align}
where $D=\rho w$ is a conserved mass density. $D_{\rm tol}$, $D_{\rm rel}$, and $D_{\rm atm}$ are chosen such that the electric conductivity has a smooth transition from high conductivity, i.e., ideal MHD, in a high-density region to zero conductivity in a low-density region. 
Specifically, they chosed $D_{\rm tol}=0.01$, $D_{\rm rel}=100\rho_{\rm atm}$, $D_{\rm atm}=\rho_{\rm atm}$, and $\sigma_0=2\times 10^{11}~{\rm s^{-1}}$. $\rho_{\rm atm}$ is a tenuous atmosphere outside the neutron star and set to be $6.17\times 10^{6}~{\rm g~cm^{-3}}$. The pre-merger magnetic field is $1.97\times 10^{12}~{\rm G}$. The employed grid resolution is $\Delta x =148~{\rm m}$. The neutron star is modeled with the $\Gamma$-law equation of state with $\Gamma=2$. Their main finding is that the magnetic winding inside the remnant massive neutron star for the resistive MHD case becomes inefficient compared to the ideal MHD case because the magnetic field is not perfectly frozen into the fluid elements. As a result, the lifetime of the remnant massive neutron star becomes $\sim 2~{\rm ms}$ longer in the resistive MHD case, which results in more massive torus formation after the remnant collapses into a black hole. However, these resistive effects need to be more clarified because the magnetic braking timescale with $\sim 10^{12}~{\rm G}$ is much longer than the lifetime of their particular model. 

Palenzuela and his collaborators performed magnetized binary neutron star merger simulations incorporating neutrino physics for the first time~\cite{Palenzuela:2015dqa}. 
They implemented the finite temperature nuclear density equation of states, the SFHo~\cite{Steiner:2012rk}, DD2~\cite{Hempel:2009mc}, and NL3~\cite{Hempel:2009mc}, and the neutrino leakage to take into account the neutrino cooling~\cite{OConnor:2009iuz}. The former results in a short-lived merger remnant, and the latter two do in a long-lived merger remnant. They also employed a sub-grid model similar to Eqs.~(\ref{eq:chapter16_subgrid})--(\ref{eq:chapter16_subgrid2}) to capture the unresolved Kelvin-Helmholtz instability. The pre-merger magnetic field strength is $10^{13}~{\rm G}$, and the grid resolution is $\Delta x=230~{\rm m}$.Their findings are summarized as follows: (i) The magnetic field is exponentially amplified up to the saturation energy of $\sim 10^{50}~{\rm erg}$. (ii) The post-merger evolution of the remnant massive neutron star is not sensitive to the amplified magnetic field, at least within the timescale of $\approx 10~{\rm ms}$ after the merger. (iii) The angular momentum transport seems to be facilitated by the magnetic braking due to the amplified magnetic field. However, there are a couple of caveats. The first one is the sub-grid term breaks the divergence-free condition since their MHD implementation is based on ``${\bf B}$," not on ``${\bf A}$." The second one is the saturation energy of the magnetic field is very sensitive to the sub-grid implementation since the saturation energy is one order magnitude smaller than that reported in Ref.~\cite{Giacomazzo:2014qba}. The third point is the role of the MRI turbulence for the angular momentum transport needs to be more clarified since they did not analyze the MRI. 

\begin{figure}[b]
\includegraphics[scale=.75]{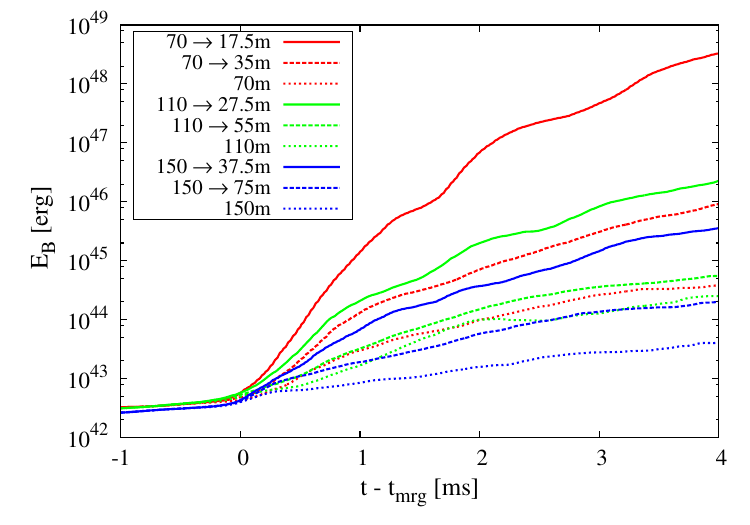}
\includegraphics[scale=.75]{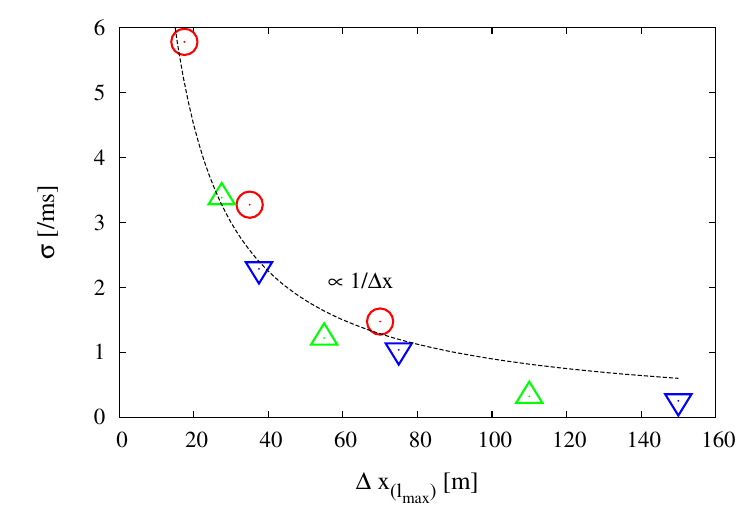}

\caption{(Top) Magnetic-field energy as a function of the post-merger time with the pre-merger magnetic field of $10^{13}~{\rm G}$. The legend denotes the employed grid resolution. For example, the red curves correspond to the base run with $\Delta x = 70~{\rm m}$ in the dashed curve, the run with $\Delta x = 35{\rm m}$ by the increment of one mesh refinement domain from the base run in the dotted curve, and the run with $\Delta x = 17.5~{\rm m}$ by the increment of two mesh refinement domains from the base run in the solid curve. (Bottom) The measured growth rate of the magnetic field energy as a function of the employed grid resolution. The color scheme is the same as the upper panel. 
The figures are taken from Ref.~\cite{Kiuchi:2015sga}.  }
\label{fig:chapter16_Kiuchi2015a} 
\end{figure}

\begin{figure}[b]
\includegraphics[scale=.75]{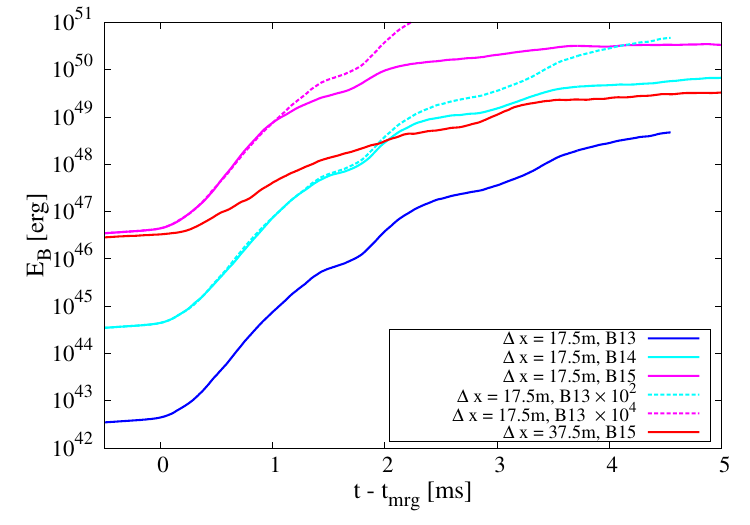}
\caption{Magnetic-field energy as a function of the post-merger time. B13, B14, and B15 denote the pre-merger magnetic field of $10^{13}~{\rm G}$, $10^{14}~{\rm G}$, and $10^{15}~{\rm G}$, respectively. The cyan and purple dashed curves are the results with $\Delta x=17.5~{\rm m}$ and the pre-merger magnetic field of $10^{13}~{\rm G}$ magnified by a factor of $10^2$ and $10^4$, respectively.
The figures are taken from Ref.~\cite{Kiuchi:2015sga}. }
\label{fig:chapter16_Kiuchi2015b} 
\end{figure}

Kiuchi and his collaborators reported new magnetized binary neutron star merger simulations~\cite{Kiuchi:2015sga}. The pre-merger magnetic fields are $10^{13}~{\rm G}$, $10^{14}~{\rm G}$, and $10^{15}~{\rm G}$. The employed grid resolutions are $\Delta x=17.5~{\rm m}$, $27.5~{\rm m}$, and $37.5~{\rm m}$ for the convergence test. It should be noted that the high-resolution mesh refinement domains are assigned only to a region where the shear layer appears. The entire neutron stars in the pre-merger phase were covered by $70~{\rm m}$, $110~{\rm m}$, and $150~{\rm m}$, respectively. The NS model is the same as that in Ref.~\cite{Kiuchi:2014hja}. They measured the growth rate of the magnetic field amplification due to the Kelvin-Helmholtz instability, and it is approximately proportional to the inverse of the employed grid resolution (see Fig~\ref{fig:chapter16_Kiuchi2015a}). They also explored the saturation energy of the magnetic field by varying the pre-merger magnetic field. The back reaction due to the amplified magnetic field is likely to start once the magnetic-field energy reaches $\sim 10^{49}~{\rm erg}$, and it is likely to saturate at $\sim 10^{50}~{\rm erg}$ (see Fig~\ref{fig:chapter16_Kiuchi2015b}). However, the energy spectrum analysis indicated that only a small fraction of $\sim 0.5$--$1$ \% of the turbulent kinetic energy is transferred to the magnetic field energy. Also, there is no prominent sign of the angular momentum transport at $\approx 5~{\rm ms}$ after the merger because, for efficient magnetic braking, a large-scale poloidal magnetic field is necessary. It implies that (i) the physical saturation of the magnetic field energy could still be far from that reported in the simulations, and (ii) the Kelvin-Helmholtz instability only produces the small-scale field. The MRI analysis was missed in their work because of the shortness of the simulation, i.e., $5~{\rm ms}$ after the merger. 

\begin{figure}
  \centering
  \includegraphics[width=0.75\textwidth]{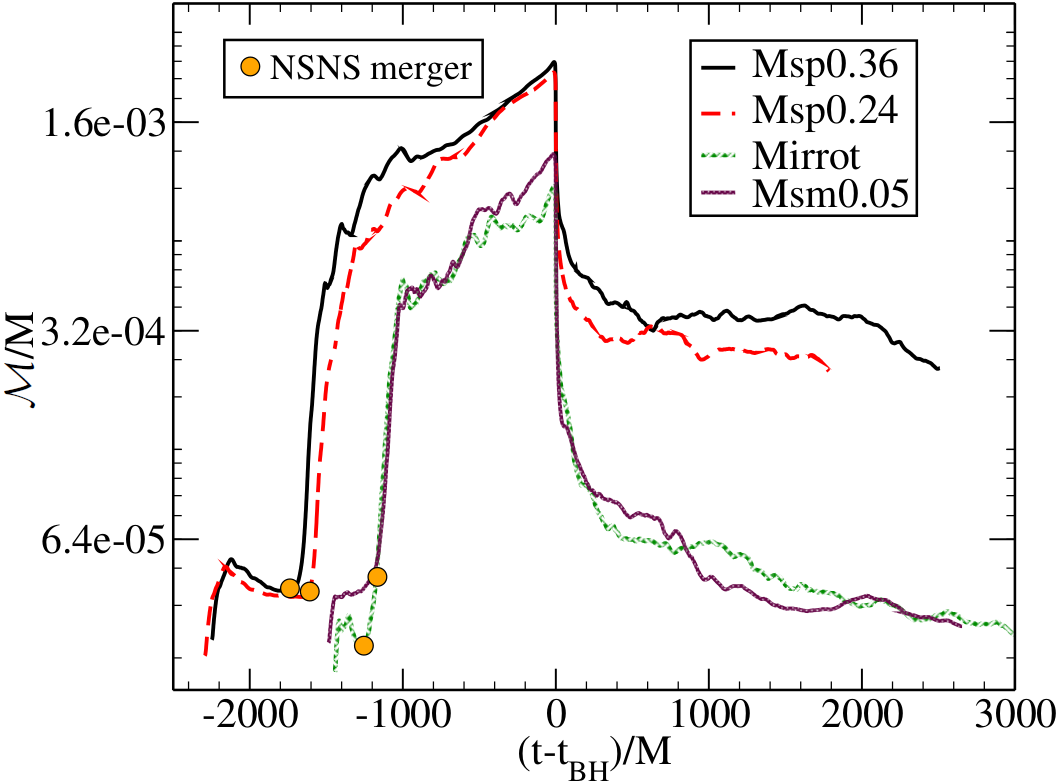}
  \caption{Total magnetic energy $\mathcal{M}$ normalized by the ADM mass
    $M=5.36\times 10^{54}(M_{\rm NS}/1.625M_{\odot})\rm erg$ versus time. Dots indicate the
    NSNS merger time $t_{\rm mer}$. The coordinate time has been shifted to
    the BH formation time $t_{\rm BH}$. Msp0.36, Msp0.24, Mirrot, and Msm0.05 denote the magnetized binary with $\chi_{\rm NS}=0.36$, $0.24$, $0$, and $-0.05$, respectively. The figure is taken from Ref.~\cite{Ruiz:2019ezy}. 
    \label{fig:chapter16_Ruiz2019a}}
\end{figure}

Ruiz and his collaborators performed magnetized binary neutron star merger simulations~\cite{Ruiz:2016rai}. They employed $\Delta x=152~{\rm m}$ and $227~{\rm m}$. The neutron star is modeled by the $\Gamma$-law equation of state with $\Gamma=2$. The merger remnant is short-lived. The pre-merger magnetic field strength was unclear because they only mentioned the field strength at pole $B_\mathrm{pole}=1.75\times 10^{15}~{\rm G}$. However, from Fig.~5 in Ref.~\cite{Ruiz:2020via} reported by the same authors with a similar set-up except for the neutron star spin, the pre-merger magnetic field is guessed to be $\approx 10^{16.5}$--$10^{17}~{\rm G}$ (see also Fig.~\ref{fig:chapter16_Ruiz2019a}). In addition to the confined pre-merger magnetic field configuration, they employed a pulsar-like configuration:
\begin{align}
A_\phi=\frac{\pi r_0^2 I_0 \varpi^2}{(r_0^2+r^2)^{3/2}} \left(1+\frac{15r_0^2(r_0^2+\varpi^2)}{8(r_0^2+r^2)^2}\right), \label{eq:chapter16_dipole}
\end{align}
where $r_0$ is the current loop radius and $I_0$ is the loop current, $r^2=(x-x_{\rm NS})^2+(y-y_{\rm NS})^2+z^2$, and $\varpi^2=(x-x_{\rm NS})^2+(y-y_{\rm NS})^2$ where $x_{\rm NS}$ and $y_{\rm NS}$ denote the initial coordinates of the neutron star center of mass. Importantly, the neutron star exterior condition for the hydrodynamic variables is chosen such that the exterior plasma beta is $0.01$. They found a mildly relativistic outflow, an incipient jet, from the merger remnant composed of the rapidly spinning black hole and massive torus. They reported that this is the case irrespective of the pre-merger magnetic field geometry.
Since they mentioned the MRI quality factor in the remnant massive neutron star is greater than 10, Eq.~(\ref{eq:chaper16_MRI}) and their grid-set up suggested $B_p \gtrsim 3\times 10^{16}~{\rm G}$. The previous works showed that it is impossible to reproduce the efficient Kelvin-Helmholtz instability with this resolution~\cite{Giacomazzo:2010bx,Kiuchi:2014hja,Kiuchi:2015sga,Rezzolla:2011da} unless the sub-grid model is employed. Therefore, the poloidal magnetic field inside the merger remnant could be a consequence of the pre-merger large-scale magnetic field. 
After the remnant neutron star collapsed into the black hole, the pole region evacuated, and the incipient jet was launched at $\approx 40~{\rm ms}$ after the black hole formation. 

They argued the discrepancy from Ref.~\cite{Kiuchi:2014hja}, i.e., the absence/presence of the jet launching, may originate from the difference of the neutron star model. When the two neutron stars collide, the dynamical ejecta is driven by the shock heating and tidal force~\cite{Hotokezaka:2012ze}. A part of the dynamical ejecta falls back onto the merger remnant, and the resultant ram pressure may overcome the magnetic pressure. However, the fallback timescale may be different in different binary neutron star models. Namely, the fallback matter in their model is smaller than that in Ref.~\cite{Kiuchi:2014hja}. In the latter model, it may take time to launch a relativistic jet. A quantitative comparison on the fallback timescale should be necessary among the different binary neutron star merger models. Also, the large-scale magnetic field generation mechanism should be investigated in detail. 
The final caveat is the magnetic winding timescale originating from the pre-merger large-scale field is as short as $O(1)~{\rm ms}$ (see Eq.~(\ref{eq:chapter16_winding})), which is shorter than the lifetime of the remnant massive neutron star of $O(10)~{\rm ms}$. 

Endrizzi and his collaborators reported a new series of magnetized binary neutron star merger simulations~\cite{Endrizzi:2016kkf}. The pre-merger magnetic field is $10^{13}~{\rm G}$. The employed grid resolution is $\Delta x=222~{\rm m}$. The neutron star is modeled with the APR equation of state~\cite{Akmal:1998cf}. They employed a hybrid equation of state to take into account the thermal effect during the simulation. They simulated three cases: a high mass case, which results in a prompt black hole formation, and a symmetric and asymmetric low mass case, which results in a supramassive neutron star formation~\footnote{The supramassive neutron star is defined as the mass below the uniformly rotating neutron star's maximum mass and above the TOV star's maximum mass.}. 
They reported an exponential amplification of the magnetic field energy after the merger in the two later cases (see Fig.~9 in Ref.~\cite{Endrizzi:2016kkf}.) However, the density-weighted average of the magnetic field strength did not exhibit such an exponential growth (see Fig.~11 in Ref.~\cite{Endrizzi:2016kkf}.) Also, the saturation energy of the magnetic field of $10^{48}$--$10^{49}~{\rm erg}$ is much smaller than the rotational kinetic energy of $\sim 10^{53}~{\rm erg}$. As they reported, the Kelvin-Helmholtz instability and MRI are not resolved. Therefore, the detailed analysis is still necessary.

Kawamura and his collaborators reported a systematic study of magnetized binary neutron star merger simulations~\cite{Kawamura:2016nmk}. The grid resolution is $\Delta x=222~{\rm m}$ for the $\Gamma$-law equation of state with $\Gamma=2$ and $\Delta x =186~{\rm m}$ for the H4 equation of state~\cite{Glendenning:1991es}. The pre-merger magnetic field is $\sim 10^{12}~{\rm G}$ and explored the pre-merger magnetic field topology with the field aligned to the orbital rotation axis, aligned and anti-aligned field, and both anti-aligned field. They also performed the convergence test by changing the grid resolution $\Delta x=177~{\rm m}$ and $277~{\rm m}$ for the $\Gamma$-law equation of state and $\Delta x =150~{\rm m}$ for the H4 equation of state. Their results with the standard grid set-up are summarized as follows: (i) For the $\Gamma$-law equation of state, the symmetric binary results in a short-lived remnant massive neutron star formation and subsequent massive torus formation after the remnant collapses into a black hole. The Kelvin-Helmholtz instability is inefficient, and the MRI cannot be resolved during the remnant massive neutron star phase due to their set-up for the pre-merger magnetic field and grid (see Eq.~(\ref{eq:chapter16_MRI_Qfac})). Inside the massive torus, the magnetic field is amplified, but its density-weighted mean value saturated at $\approx 10^{12}~{\rm G}$. $\lambda_{\rm MRI}$ in this phase is likely to be
\begin{align}
\lambda_{\rm MRI} = \frac{B_{\rm mean}}{\sqrt{4\pi\rho}} 2\pi\left(\frac{GM_{\rm BH}}{R^3_{\rm disk}}\right)^{-1/2}\approx 10^{3}~{\rm cm}~B_{{\rm mean},12} \rho_{12}^{-1/2} M_{\rm BH,3}^{-1/2}R_{{\rm disk},6.7}^{3/2},
\end{align}
where $B_{{\rm mean},12}=B_{\rm mean}/10^{12}~{\rm G}$, $\rho_{12}=\rho/10^{12}~{\rm g~cm^{-3}}$, $R_{{\rm disk},6.7}=R_{\rm disk}/5\times10^{6}~{\rm cm}$, $M_{\rm BH,3}=M_{\rm BH}/3M_\odot$, and we assumed the Newtonian Keplarian angular velocity with the disk radius $R_{\rm disk}$. Therefore, the magnetic field amplification inside the torus is likely due to magnetic winding. However, the saturation energy of the magnetic field of $\sim 10^{44}~{\rm erg}$ is far from the equipartition to the rotational kinetic energy of $\sim 10^{50}~{\rm erg}$. Also, they reported the magnetic field amplification in each phase is very sensitive to the grid resolution. For example, the magnetic field energy differs by six orders of magnitude at the end of the simulation with high- and low-resolution runs. (ii) For the asymmetric binary case with the $\Gamma$-law equation of state, the merger remnant almost promptly collapses into a black hole, and a torus more massive than that in the symmetric binary is formed. However, the magnetic field amplification inside the massive torus is minor, i.e., only a factor of few. It is counterintuitive because the rotational energy should be tapped into the amplification of the magnetic field energy via the MRI and magnetic winding, i.e., the more massive torus is expected to have stronger magnetization. 
(iii) For the symmetric binary with the H4 equation of state, the magnetic field evolution until the black hole formation is qualitatively similar to the symmetric binary with the $\Gamma$-law equation of state. However, the magnetic field is not amplified inside the massive torus and stays at $\sim 10^{44}~{\rm erg}$ until the end of the simulation. It is a striking difference between their result and Ref.~\cite{Kiuchi:2014hja} reporting the saturation energy is likely to be $\sim 10^{49}~{\rm erg}$ (see also Fig.~\ref{fig:chapter16_Kiuchi2014}) with the same binary, but different pre-merger magnetic field and grid set-up. They performed the convergence study with $\Delta x=150~{\rm m}$ and reported a significant growth of the magnetic field during the massive torus phase, presumably due to the non-axisymmetric MRI, which is not observed in their standard resolution. In the high-resolution run, the magnetic field energy is amplified up to $\sim 10^{50}~{\rm erg}$ inside the torus at the end of the simulation. 
(iv) For the asymmetric binary with the H4 equation of state, they reported an exponential growth of the magnetic field during a remnant massive neutron star phase. However, given the density-weighted mean value of the magnetic field of $\sim 10^{12}$--$10^{13}~{\rm G}$ (see Fig.~13 in Ref.~\cite{Kawamura:2016nmk}), the MRI is likely to be unresolved~(\ref{eq:chapter16_MRI_Qfac}). As with Ref.~\cite{Endrizzi:2016kkf}, the origin of the exponential growth is a riddle. 
It should be noted that the MRI quality factor estimated in Ref.~\cite{Kiuchi:2014hja} and this work is {\it not} based on the poloidal field strength, i.e., $B_{\rm p}$ in Eq.~(\ref{eq:chaper16_MRI}), but the total magnetic field strength, implying the most likely to be the toroidal field strength. In other words, they explored the non-axisymmetric MRI, not the axisymmetric MRI~\cite{Balbus:1991ay}. Since the axisymmetric MRI has a more significant growth rate than the non-axisymmetric MRI~\cite{Balbus:1998ja}, and the poloidal field is generally weaker than the toroidal field, resolving the axisymmetric MRI is more challenging. 

Ciolfi and his collaborators reported new magnetized binary neutron star merger simulations, which leave a long-lived remnant~\cite{Ciolfi:2017uak}. They employed the grid resolution of $\Delta x=222~{\rm m}$ and $177~{\rm m}$ for the convergence test. The neutron star is modeled with the APR equation of state~\cite{Akmal:1998cf}, the MS1 equation of state~\cite{Mueller:1996pm}, and the H4 equation of state~\cite{Glendenning:1991es}. They explored the symmetric and asymmetric binaries. The pre-merger magnetic field is $\approx 3\times 10^{15}~{\rm G}$. They reported that after an inefficient Kelvin-Helmholtz instability amplification, the MRI is likely to play an active role in their models. However, there are a couple of caveats: (i) Since $\lambda_{\rm MRI}$ is not estimated by the poloidal field strength but by the presumably dominant toroidal field strength, it seems to be overestimated as they mentioned by themselves. (ii) Since they have not quantified the MRI quality factor with a number, it is hard to judge the ability of their simulation setup to resolve the MRI. As shown in independent simulations, it seems to be hard to resolve the axisymmetric MRI with their setup~\cite {Kiuchi:2017zzg,Kiuchi:2023obe} (see below). 

Ruiz and his collaborators reported magnetized binary neutron star mergers in a prompt black hole formation scenario in Ref.~\cite{Ruiz:2017inq}. 
The grid resolution is $\Delta x\approx 140$--$150~{\rm m}$. 
The pre-merger magnetic field is the same as in their previous work~\cite{Ruiz:2016rai}, i.e., likely to be $10^{16.5}$--$10^{17}~{\rm G}$ (see also their Table I which shows the pre-merger magnetic field energy is as high as $10^{50}~{\rm erg}$). In additon to the confined pre-merger mangetic field configuration, they explored the dipole-like field configuration (see Eq.~(\ref{eq:chapter16_dipole})). 
They confirmed the MRI is resolved inside the massive torus around a black hole and estimated the Shakura-Sunyaev parameter~\cite{Shakura:1973} for the first time. However, contrary to the remnant massive neutron star formation case~\cite{Ruiz:2017inq}, they did not find a launching of an incipient jet irrespective of the mass asymmetry of the binary neutron star. 
Since the shocked component of the dynamical ejecta and resultant fallback matter in the polar region should be negligible in the prompt black hole formation, the conclusion is puzzling compared to their previous work, i.e., the less fallback matter is favorable for the launching of the incipient jet. The absence of a large-scale magnetic field inside the disk needs to be clarified. 

Kiuchi and his collaborators reported new magnetized neutron star merger simulations. The employed grid resolution is $\Delta x =12.5~{\rm m}$~\cite{Kiuchi:2017zzg}. Similar to their previous work~\cite{Kiuchi:2015sga}, the high-resolution mesh refinement domains are assigned to a shear layer, and the entire binary neutron star before the merger is covered by $\Delta x=50~{\rm m}$. The pre-merger magnetic field is $10^{15}~{\rm G}$. The neutron star is modeled with the H4 equation of state~\cite{Glendenning:1991es}. They also performed the convergence study with $\Delta x=70~{\rm m}$ and $110~{\rm m}$. They explored the less massive binary neutron star merger compared to that in Refs.~\cite{Kiuchi:2014hja,Kiuchi:2015sga}, resulting in the long-lived massive neutron star. They thoroughly assessed the MRI-driven turbulence in a non-linear phase with the axisymmetric, the non-axisymmetric MRI quality factor, the ratio of the poloidal field energy to the toroidal field energy, the Maxwell stress, and associated Shakura-Sunyaev parameters~\cite{Hawley:2011tq}. 
Their findings are summarized as follows: (i) The MRI quality factor, as well as the other diagnostics to quantify the ability of the simulation to sustain the MRI-driven turbulence, indicates that the resolution coarser than $\Delta x=70~{\rm m}$ is insufficient in the region of $\rho \ge 10^{13}~{\rm g~cm^{-3}}$. Particularly, with $\Delta x =110~{\rm m}$, which is a ``high" resolution in the numerical relativity community, the MRI-driven turbulence cannot be sustained in the region $\rho \ge 10^{12}~{\rm g~cm^{-3}}$ (see Table I in Ref.~\cite{Kiuchi:2017zzg}). It should be noted that a bulk of the rotational energy of the remnant massive neutron star is contained in such a high-density region. Unresolved MRI simulation could draw a physically {\it incorrect} picture unless the MRI is suppressed by a mechanism such as neutrino viscosity and drag. (ii) An ordered (large-scale) poloidal magnetic field is not established within their simulation timescale of $\approx 30~{\rm ms}$ after the merger. It indicates the magnetic braking timescale could be $\approx 0.7$--$0.8~{\rm s}$ unless the large-scale poloidal field is not established by a mechanism such as $\alpha\Omega$ dynamo. The absence of a large-scale field needs to be clarified. 

Ruiz and his collaborators placed an upper limit of the maximum mass of the TOV star by combining their simulation results~\cite{Ruiz:2016rai,Ruiz:2017inq} and GW170817 observation associated with GRB170817A~\cite{Goldstein:2017mmi,Ruiz:2017due}. In addition to the results in Refs.~\cite{Ruiz:2016rai,Ruiz:2017inq}, they performed a simulation for a supramassive neutron star formation case, i.e., the lifetime of the merger remnant could be longer than the magnetic dipole radiation timescale. They concluded that the supramassive neutron star can not launch the jet. Based on these results, their scenario is that the merger remnant in GW170817 is between the supramassive limit and the prompt collapse limit. Since we know the mass of the binary neutron star in GW170817, this scenario gives 
\begin{align}
M^{\rm sup}_{\rm max} \approx \beta M^{\rm TOV}_{\rm max} \le M_{GW170817} \approx 2.74M_\odot \le \alpha M^{\rm TOV}_{\rm max},
\end{align}
where $\alpha=1.3$--$1.7$ determines the threshold mass for the prompt black hole formation~\cite{Bauswein:2013jpa,Bauswein:2017vtn,Hotokezaka:2011dh,Kashyap:2021wzs,Kolsch:2021lub,Koppel:2019pys,Tootle:2021umi}, and $M^{\rm TOV}_{\rm max}$ is the maximum mass of the TOV star, and $M^{\rm sup}_{\rm max}$ is the maximum mass of the uniformly rotating star. A Rhoades-Ruffini causality argument on the equation of state gives $\beta \approx 1.27$~\cite{Rhoades:1974}. Equilibrium configurations of uniformly rotating neutron stars with various kinds of equations of states give $\beta \approx 1.2$~\cite{Breu:2016ufb,Cook:1993qr,Lasota:1995eu,Musolino:2023edi}. 
Therefore, the upper limit of the maximum mass of the TOV star is likely to be $\approx 2.16$--$2.28M_\odot$. The caveat is that their upper limit estimation of the TOV maximum mass is based on the hypothesis that the supramassive neutron star cannot launch a relativistic jet. 

\begin{figure*}
  \centering
  \includegraphics[width=0.49\textwidth]{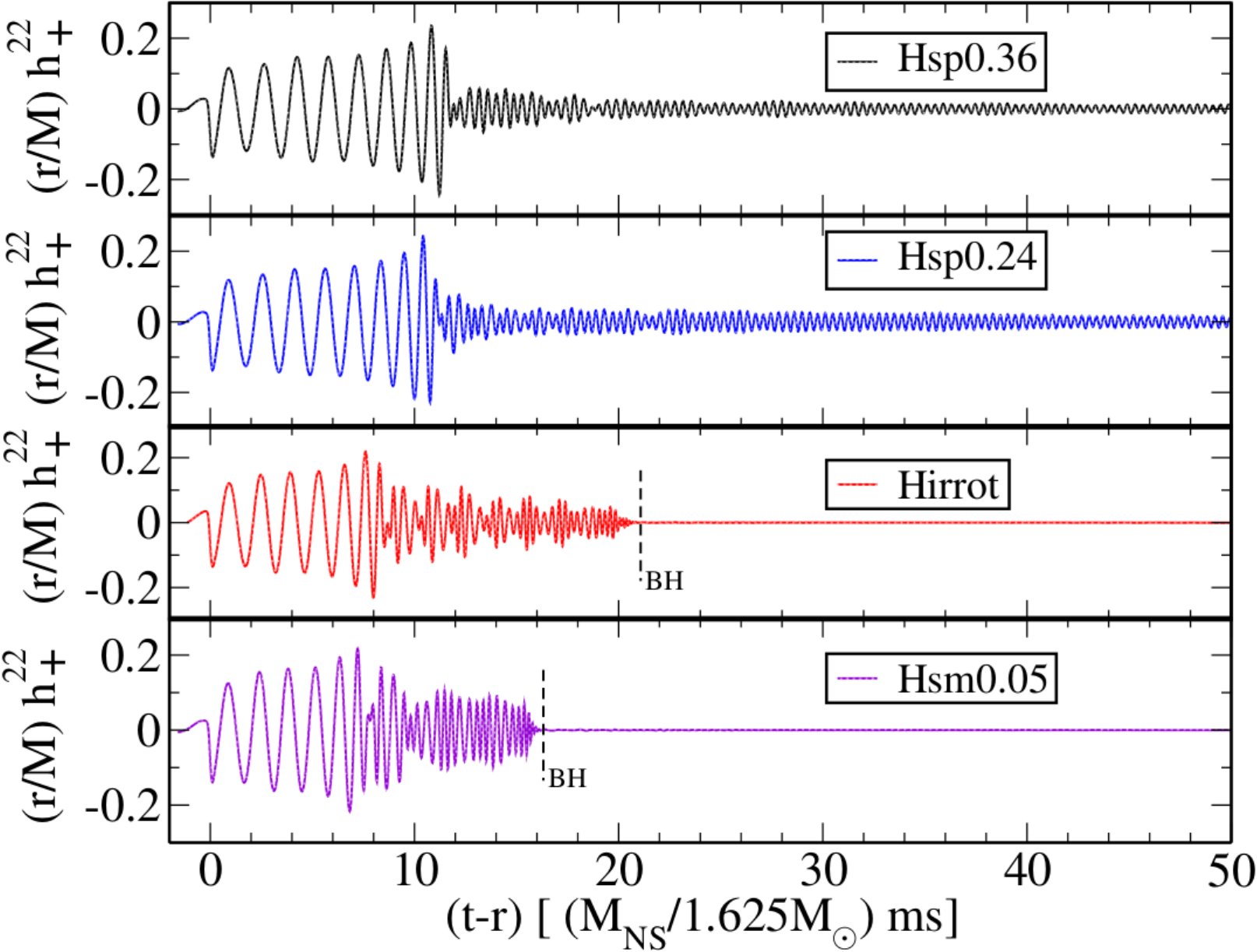}
    \includegraphics[width=0.49\textwidth]{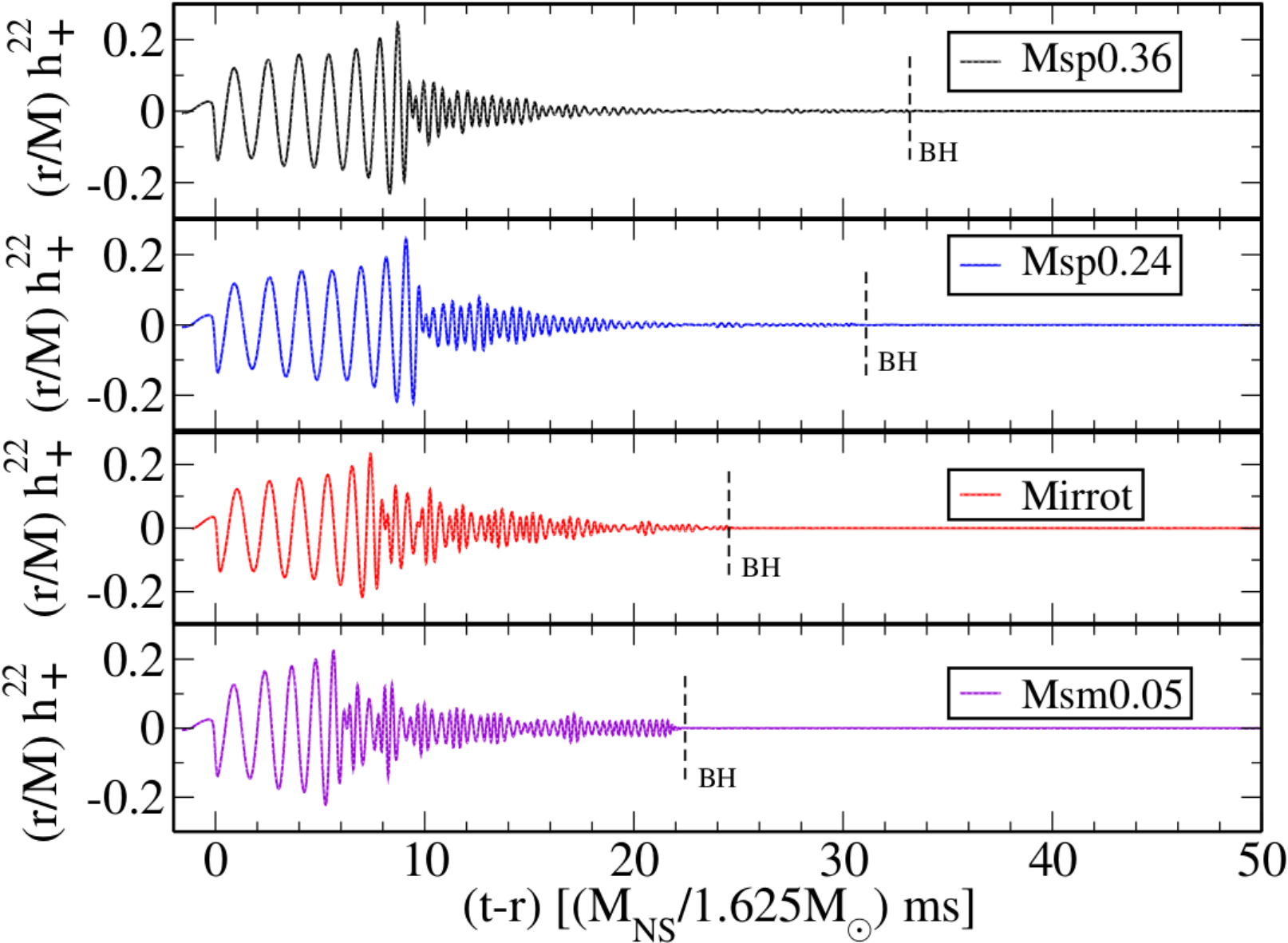}
    \caption{GW strain $h_+^{22}$ (dominant mode) as functions of retarded time. The left panel displays the GW strain
      in the unmagnetized cases, while the right panel displays the magnetized cases.
      The dashed vertical line denotes the BH formation time. Hsp0.36, Hsp0.24, Hirrot, and Hsm0.05 denote the unmagnetized binary with $\chi_{\rm NS}=0.36$, $0.24$, $0$, and $-0.05$, respectively.
      The figure is taken from Ref.~\cite{Ruiz:2019ezy}. 
    \label{fig:chapter16_Ruiz2019b}}
\end{figure*}

Ruiz and his collaborators explored a neutron star spin effect in magnetized binary neutron star mergers~\cite{Ruiz:2019ezy}. The grid resolution is $\Delta x =220~{\rm m}$. The neutron star is modeled with $\Gamma$-law equation of state with $\Gamma=2$. The pre-merger magnetic field is likely to be $10^{16.5}$--$10^{17}~{\rm G}$ with pulsar-like dipole magnetic field topology (see Eq.~(\ref{eq:chapter16_dipole})). 
Non-dimensional neutron star spin $\chi_{\rm NS}$ is $-0.05$, $0.24$, and $0.36$~\footnote{Minus sign denotes an anti-align spin with respect to the orbital angular momentum.}.
It should be noted that the fastest spin observed in the binary pulsars is $\chi_{\rm NS}\sim 0.05$~\cite{Lorimer:2008se}.
They simulated both unmagnetized and magnetized binaries. Figure~\ref{fig:chapter16_Ruiz2019a} shows the magnetic field energy, and it shows the pre-merger magnetic field energy of $\sim 2$--$3\times 10^{50}~{\rm erg}$ is amplified by a factor of $\approx 10$--$15$ at $\approx 2~{\rm ms}$ after the merger. Subsequently, the magnetic field energy continues to be amplified to $\sim 10^{52}~{\rm erg}$ until the black hole formation. They analyzed the MRI quality factor and Shakura-Sunyaev parameter. They concluded that the magneto-turbulent state is established and, consequently, the angular momentum transport is facilitated.
Figure~\ref{fig:chapter16_Ruiz2019b} shows the gravitational waves in all the models, which summarizes the unmagnetized and magnetized binary evolution. For the aligned high spinning cases with $\chi_{\rm NS}=0.36$ and $0.24$, on the one hand, the unmagnetized binaries do not collapse into a black hole until the end of the simulation. On the other hand, the magnetized counterparts do. They argued that it indicates the MHD effect facilitates the angular momentum transport. For the non-spinning case with $\chi_{\rm NS}=0$ or the anti-aligned low spinning case with $\chi_{\rm NS}=-0.05$, the magnetized binaries survive longer than the unmagnetized counterparts. Their interpretation for this is due to the efficient dissipation of the kinetic energy through the magneto-turbulence; the enhanced thermal pressure supports the remnant massive neutron stars. Also, they reported the incipient jet launching after the black hole formation in the magnetized binaries irrespective of the neutron star spin. 

However, there are a couple of caveats: (i) With their grid set up and the figure showing $Q_{\rm MRI}\ge 10$ (see Fig. 9 in Ref.~\cite{Ruiz:2019ezy}), the poloidal magnetic field strength inside the remnant is estimated to be $\gtrsim 10^{16.3}~{\rm G}$, which is comparable to the pre-merger magnetic field strength. The resultant winding timescale is $O(1){\rm ms}$ (see Eq.~(\ref{eq:chapter16_winding})). Therefore, the MHD effect initiated from the pre-merger field is not negligible. 
Also, the large-scale field generation mechanism from the small-scale MRI-amplified field needs to be clarified more. 
(ii) The estimated MRI-turbulent viscous timescale scale of $\sim 10~{\rm ms}$ (see Eq.~(4) in Ref.~\cite{Ruiz:2019ezy}) is ten times shorter than that in Ref.~\cite{Kiuchi:2017zzg} (see the second and third column in Table~I in Ref.~\cite{Ruiz:2019ezy}). Both works estimated the Shakura-Sunyaev parameter as $\sim 10^{-2}$. Strictly speaking, the binary neutron star model is different. However, it is unlikely to make an order-of-magnitude difference in the viscous timescale. The discrepancy needs to be clarified more. (iii) Their argument on the efficient dissipation of the kinetic energy through the MRI-driven turbulence in $\chi_{\rm NS}=0$ and $-0.05$ models compared to $\chi_{\rm NS}=0.36$ and $0.24$ models may not be supported by their estimated Shakura-Sunyaev parameters because it does not significantly differ among the models (see Table III in Ref.~\cite{Ruiz:2019ezy}). Note that the viscous heating is proportional to the viscous parameter. 

Ciolfi and his collaborators reported for the first time a $100~{\rm ms}$ simulation for long-lived magnetized binary neutron star merger remnant~\cite{Ciolfi:2019fie}. The pre-merger magnetic field is $10^{16}~{\rm G}$. The grid resolution is $\Delta x=220~{\rm m}$. The neutron star is modeled with the APR equation of state~\cite{Akmal:1998cf}. 
They claimed that after an inefficient Kelvin-Helmholtz instability amplification, the MRI starts to be resolved, and the magnetic field energy is amplified up to $\sim 10^{51}~{\rm erg}$. Also, they reported there is no sign of the jet launching at the end of the simulation of $\approx 100~{\rm ms}$, and they concluded the magnetar scenario for the short gamma-ray bursts is unlikely, or it could launch a jet much later than $100~{\rm ms}$. 

However, there are a couple of caveats: (i) As in Ref.~\cite{Ruiz:2019ezy}, the magnetic winding initiated from the pre-merger magnetic field is not negligible. (ii) With their grid setup and pre-merger magnetic field, it is hard to resolve the MRI, as they mentioned. Particularly, the increase of the MRI quality factor is likely caused by the magnetic winding, i.e., the MRI quality factor is for the toroidal field as they did previously in Refs.~\cite{Ciolfi:2017uak}. Also, Ref.~\cite{Kiuchi:2017zzg} suggested the axisymmetric MRI in the high-density region with $\rho \ge 10^{13}~{\rm g~cm^{-3}}$ can not be resolved with their setup (see Table I in Ref.~\cite{Kiuchi:2017zzg} with the factor of five boosts, i.e., the pre-merger magnetic field ten times larger and the grid resolution twice coarser compared to the run with $\Delta x = 110~{\rm m}$). 
(iii) The angular momentum transport due to the magnetic field needs to be clarified more as they mentioned because the angular velocity evolution, particularly in the core of the merger remnant, is similar to the unmagnetized case. 

Ruiz and his collaborators explored the magnetic-field topology in binary neutron star mergers~\cite{Ruiz:2020via}. The pre-merger magnetic field is an aligned dipole-like configuration to the orbital angular momentum and a perpendicular dipole-like configuration with a field strength of $\approx 10^{16.5}$--$10^{17}~{\rm G}$ (see Eq.~(\ref{eq:chapter16_dipole})). The grid resolution is $\Delta x =220~{\rm m}$. The neutron star is modeled with the $\Gamma$-law equation of state with $\Gamma=2$. 
The neutron star has a spin of $\chi_{\rm NS}=0.36$. Compared to the aligned--aligned pre-merger magnetic configuration in Ref.~\cite{Ruiz:2019ezy}, the aligned-perpendicular pre-merger magnetic configuration results in a longer lifetime of the remnant massive neutron star. They pointed out that it is caused by an inefficient angular momentum transport due to the MRI-driven turbulence because the Shakura-Sunyaev parameter is smaller than that in the aligned-aligned case. 
After the black hole formation, the incipient jet is launched as in the aligned-aligned case because the region with the magnetization parameter greater than $10$ appears above the black hole. For the perpendicular-perpendicular case, the lifetime of the remnant massive neutron star is longer than the aligned-aligned or aligned-perpendicular case due to the inefficient angular momentum transport. 
Consequently, the torus mass formed after the remnant massive collapses into a black hole is larger than that in the other cases. Also, the magnetic field energy at the collapse into the black hole is ten times smaller than the aligned-perpendicular case. They mentioned that the MRI quality factor is below $6$ in the perpendicular-perpendicular case, and the resultant turbulent viscosity quantified by the Shakura-Sunyaev parameter is small. They ran a higher resolution simulation, but the result is likely to be less conclusive because improving the grid resolution by only $25\%$ is not enough to capture the MRI. After the black hole formation, they did not find an incipient jet launch in the perpendicular-perpendicular model, even though it has a larger torus mass compared to the aligned-aligned case and aligned-perpendicular case. They concluded that it is consistent with the prompt collapse case, which suggests the magnetic field energy should be more significant than $0.1\%$ of the initial ADM mass at the formation of the black hole to launch the incipient jet. 

However, in addition to the caveats raised above, there is a caveat to explore the pre-merger magnetic field topology. As demonstrated in this work, it is more challenging to reproduce the MRI-driven turbulence in the perpendicular-perpendicular case, which is likely to be natural since the aligned case is preferable to resolve the MRI, i.e., $\lambda_{\rm MRI}\propto B_z$. It implies that it may not be a fair comparison for the different pre-merger magnetic field topologies with the {\it same grid set up} because the inefficient development of the MRI-driven turbulence may be merely a consequence of the numerics. 

Ciolfi reported a long-term magnetized binary neutron star merger simulation up to $250~{\rm ms}$ after the merger for the first time~\cite{Ciolfi:2020hgg}. 
The pre-merger magnetic field is $10^{15}~{\rm G}$ and $5\times 10^{15}~{\rm G}$. The employed grid resolution is $\Delta x=250~{\rm m}$. 
The neutron star is modeled by the APR equation of state~\cite{Akmal:1998cf}. His finding is summarized as follows: (i) The magnetically collimated outflow appears in the strongly magnetized case but not in the weakly magnetized case. (ii) The outflow is mildly relativistic with $\lesssim 0.3~{\rm c}$ and the kinetic energy of the outflow within an opening angle of $15^\circ$ is $\approx 3\times 10^{49}~{\rm erg}$. Therefore, except for a very optimistic scenario, the system is not likely to drive a relativistic jet compatible with GRB 170817A~\cite{Goldstein:2017mmi}. There are a couple of caveats: (i) The subtleness of the pre-merger magnetic field needs to be clarified more, particularly by conducting a detailed analysis of the MRI. (ii) The large-scale magnetic field generation inside the merger remnant must also be clarified more. Since the simulation set-up cannot resolve the MRI, the magnetic winding due to the pre-merger magnetic field relic could be the outflow generation mechanism. 

M\"{o}sta and his collaborators performed magnetized binary neutron star remnant simulations with neutrino physics~\cite{Mosta:2020hlh}. First, they performed an unmagnetized binary neutron star merger simulation. Then, they embeded a dipole-like seed poloidal magnetic field of $10^{15}~{\rm G}$ inside the remnant massive neutron star, which is short-lived. They took into account the neutrino cooling by the leakage scheme and the neutrino heating phenomenologically. They performed an in-depth convergence test with $\Delta x=55~{\rm m}$, $110~{\rm m}$, and $220~{\rm m}$. The neutron star is modeled with LS220~\cite{Lattimer:1991nc}. They also explored the binary merger remnant evolution without neutrino. Their findings are summarized as follows: (i) The highest resolution is fine enough to capture the MRI, and the toroidal magnetic field amplification is significantly enhanced compared to the low-resolution run. The enhanced toroidal field drives an MHD wind with faster velocity than the non-magnetized case, which is the neutrino-driven wind. (ii) The neutrino cooling helps mitigate the polar region's baryon pollution and aids the launch of the jet with the terminal Lorentz factor of $\sim 5$. (iii) The mass ejection by the MHD effect increases with the grid resolution by a factor of three. As they mentioned, the caveat is that they assumed a dynamo activity to build up the large-scale poloidal field inside the remnant massive neutron star. 

\begin{figure}
        \centering
        \includegraphics[width=0.49\textwidth]{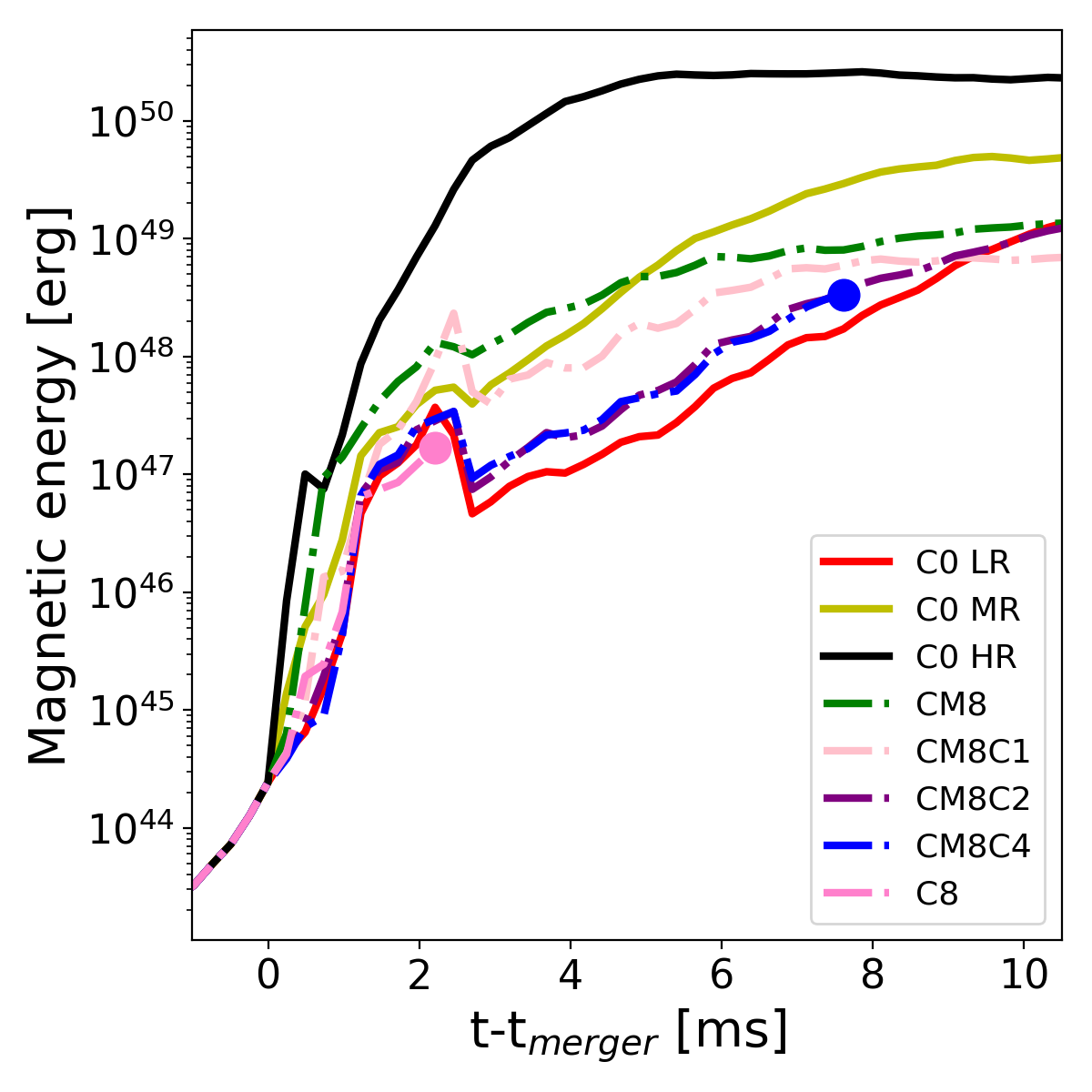}\\
        \includegraphics[width=0.49\textwidth]{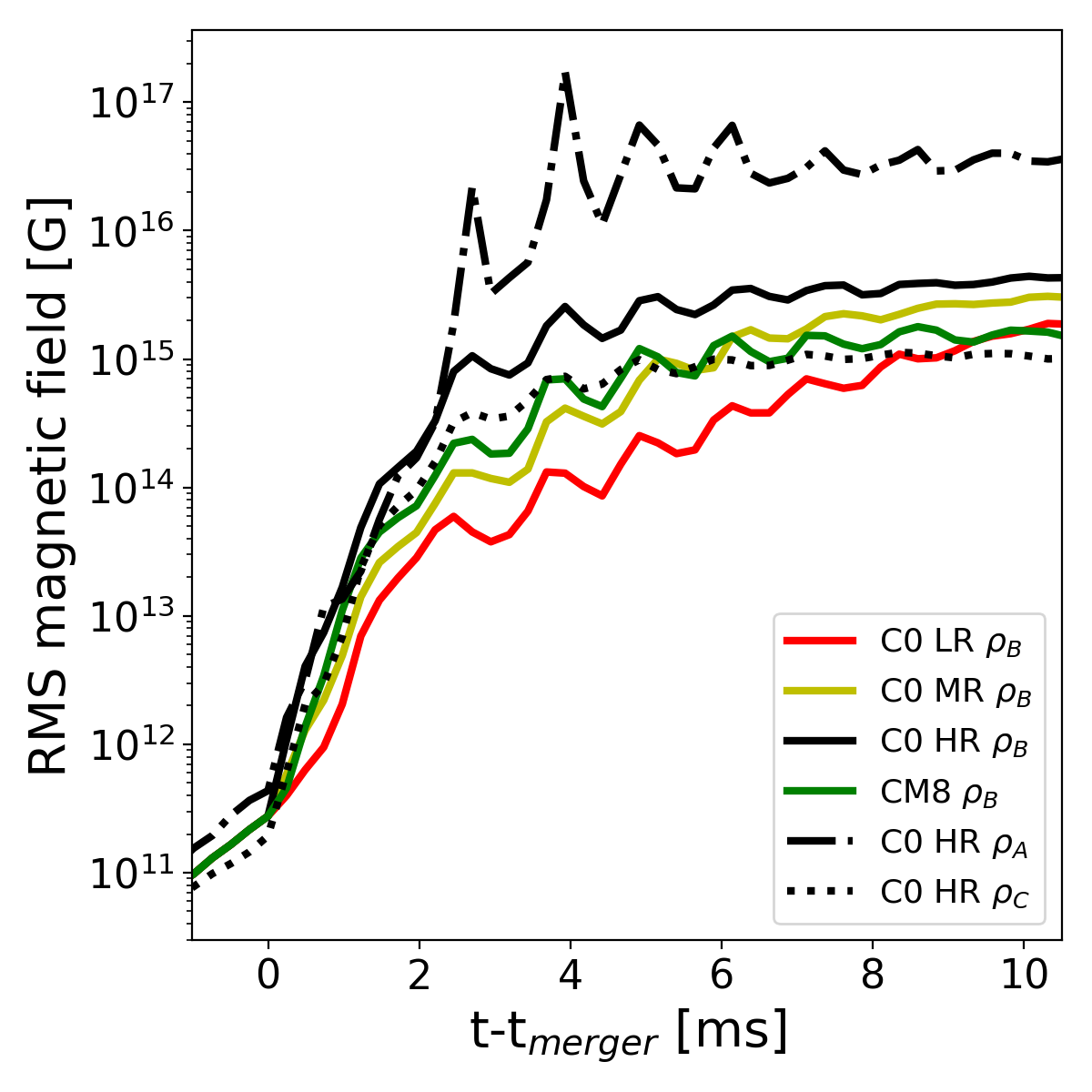}
        \caption{(Top) Integrated magnetic energy as a function of the post-merger time. The circles indicate the collapse of the remnant, forming a black hole. {\tt C0LR, C0MR}, and {\tt C0HR} in the legend denote the direct simulations with $\Delta x = 147~{\rm m}$, $74~{\rm m}$, and $37~{\rm m}$, respectively. {\tt CM8, CM8C1, CM8C2, CM8C4}, and {\tt C8} denote the sub-grid simulation with $(C_M,C_N,C_T)=(8,0,0)$, $(8,1,1)$, $(8,2,2)$, $(8,4,4)$, and $(8,8,8)$, respectively. 
        (Bottom) The root-mean-square value of the magnetic field for the direct simulations with different resolutions and for the most favorable sub-grid simulation case {\tt CM8}. The r.m.s. magnetic field of the high-resolution case, {\tt C0HR}, is calculated in different regions with $\
\rho > \rho_X~\rm{g~cm^{-3}}$ , being $\rho_A= 6 \times 10^{9}$, $\rho_B= 6 \times 10^{10}$ (value used for the top panel) and $\rho_C= 6 \times 10^{11}$. The figures are taken from Ref.~\cite{Aguilera-Miret:2020dhz}.}
        \label{fig:chapter16_Aguilera-Miret2020a}
\end{figure}

\begin{figure*}
        \centering
        \includegraphics[width=0.49\textwidth]{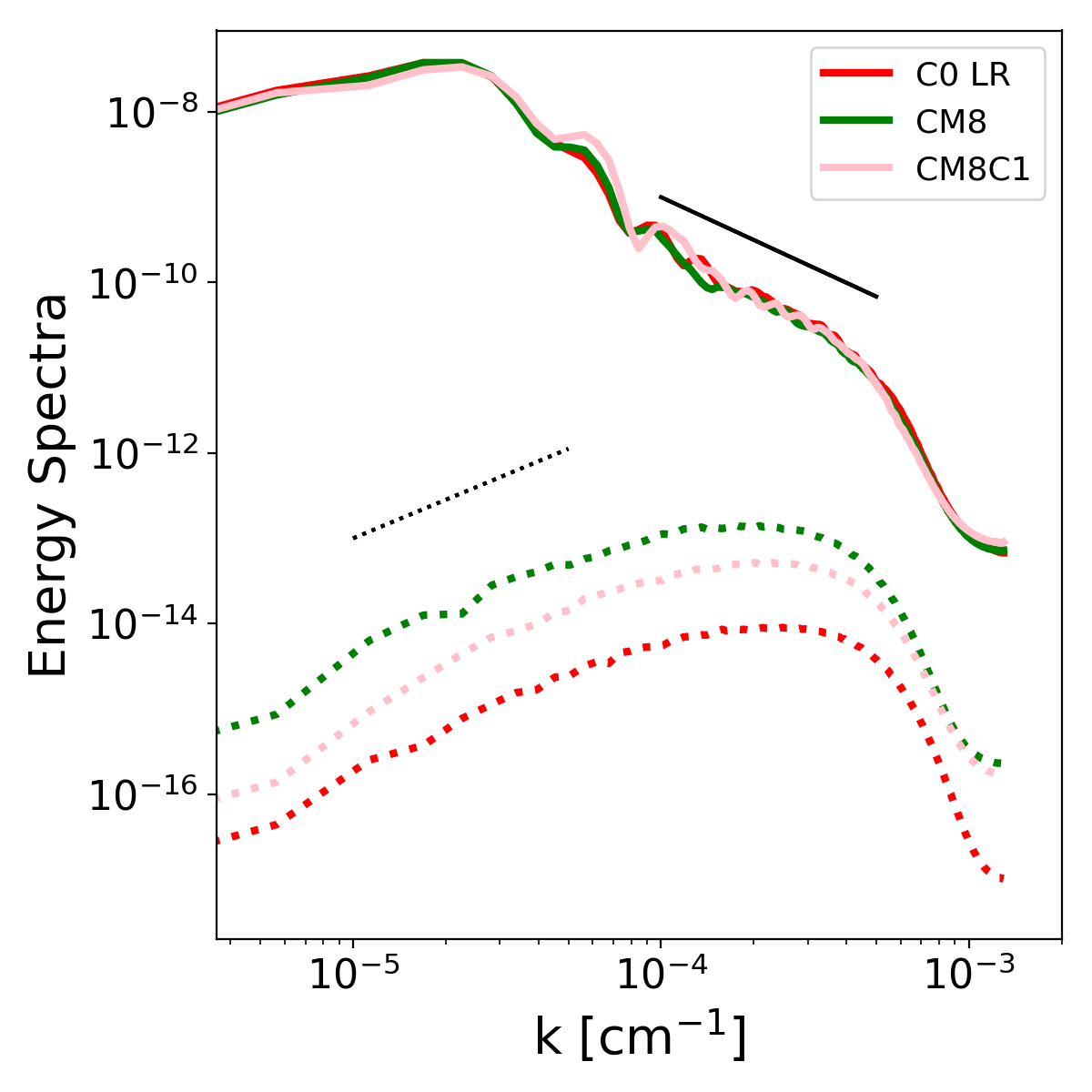}
        \includegraphics[width=0.49\textwidth]{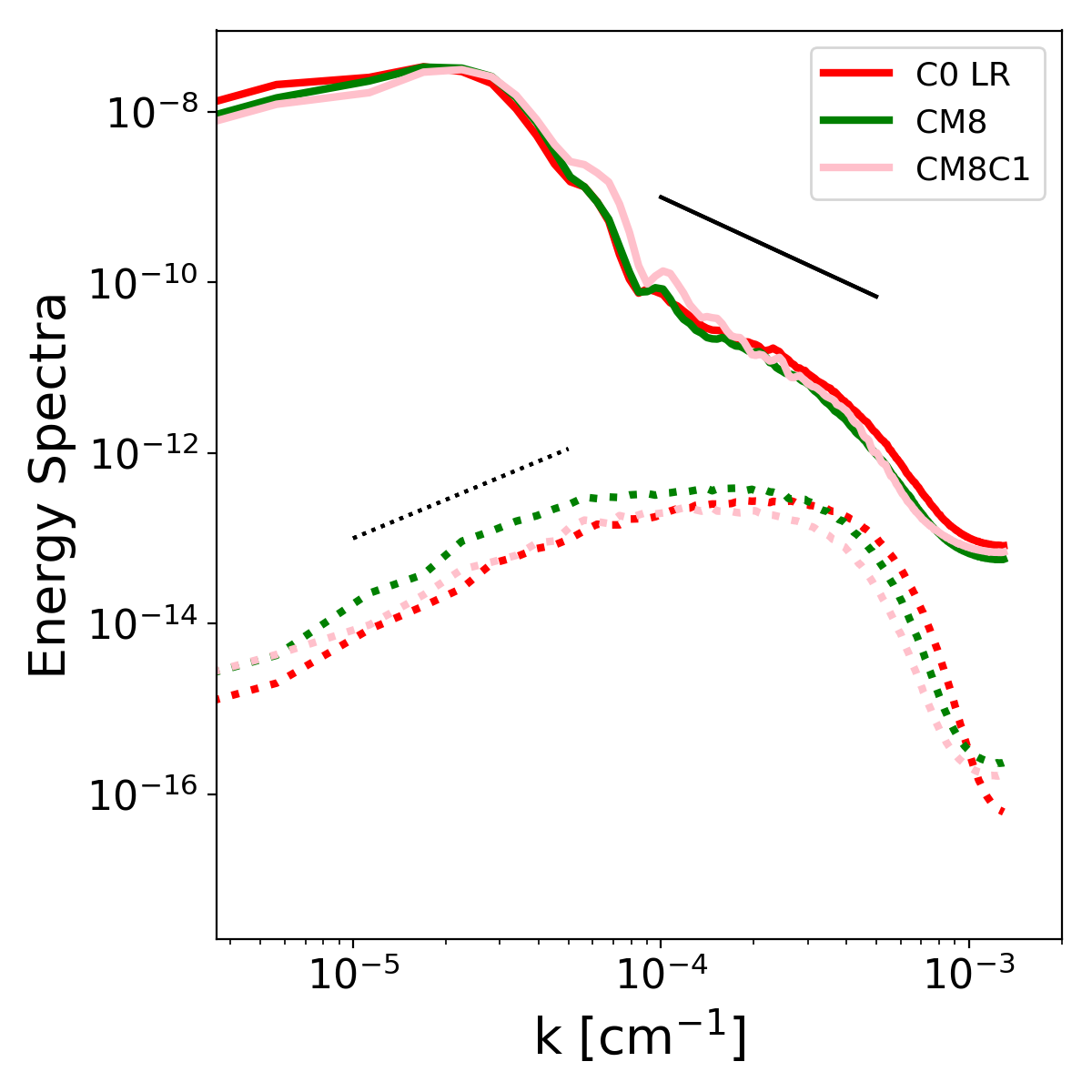}
        \caption{Magnetic (dashed) and kinetic (solid) spectra of {\tt C0LR}, {\tt CM8} and {\tt CM8C1} at $5~{\rm ms}$ (left) and $10~{\rm ms}$ (right) after the merger. Units in the vertical axis are arbitrary. Black slopes are the Kolmogorov $(\propto k^{-5/3})$ and Kazantsev $(\propto k^{3/2})$. The figure is taken from Ref.~\cite{Aguilera-Miret:2020dhz}. 
        }
        \label{fig:chapter16_Aguilera-Miret2020b}
\end{figure*}

Aguilera-Miret and his collaborators performed magnetized binary neutron star merger simulations with a sub-grid model~\cite{Aguilera-Miret:2020dhz}. Their sub-grid model is based on the gradient sub-grid scale model in which filtering with a Gaussian kernel is applied to each physical quantity~\cite{Carrasco:2019uzl}. This prescription gives additional terms in the physical flux of the MHD equation:
\begin{align}
&\tau^i_N=- C_N \xi H^i_N~\text{(the flux in the continuity equation)},\\
&\tau^{ij}_T=-C_T \xi H^{ij}_T~\text{(the flux in the momentum equation)},\\
&\tau^i_M= - C_M \xi H^i_M~\text{(fhe flux in the induction equation)} \label{eq:chpater16_SGS},
\end{align}
where $\xi=\gamma^{1/3}\Delta x^2/24$. $C_N$, $C_T$, and $C_M$ are called pre-factors, which a direct simulation should calibrate. 
The value of the pre-factors with $C_N=C_T=C_M=1$ follows from the analytical calculation of large eddy simulation with the gradient sub-grid scale model. A priori tests indicate that practically, they might slightly vary with the grid resolution and other parameters, but they usually remain between $1$--$2$~\cite{Carrasco:2019uzl}. However, the numerical dissipation inherent to the employed Riemann solver with the high-order cell reconstruction at the small scale strongly attenuates the effect of the large eddy simulation at intermediate resolutions. Practically, the authors balanced this attenuation by artificially increasing the pre-factors. 
$H^i_N$, $H^{ij}_T$, and $H^i_M$ are functions of the filtered fields, and the cumbersome expressions are found in Ref.~\cite{Carrasco:2019uzl}. 
The pre-merger magnetic field is $5\times 10^{11}~{\rm G}$, which is approximately consistent with the highly magnetized end of the binary pulsars~\cite{Lorimer:2008se}. They employed three grid resolutions of $\Delta x=37~{\rm m}$, $74~{\rm m}$, and $147~{\rm m}$ in direct simulations, i.e., without the sub-grid terms. Then, they compared them with simulations with their sub-grid model while keeping the grid resolution $\Delta x=147~{\rm m}$ and changing the value of $C_N$, $C_T$, and $C_M$. The neutron star is modeled with the SLy equation of state~\cite{Douchin:2001sv}. Figure~\ref{fig:chapter16_Aguilera-Miret2020a} summarizes their results. In the direct simulation with $\Delta x=37~{\rm m}$, the magnetic-field energy is amplified from $\sim 10^{44}~{\rm erg}$ up to $\sim 10^{50}~{\rm erg}$ due to the Kelvin-Helmholtz instability within $\approx 10~{\rm ms}$ after the merger. The growth rate of the amplification is determined by the employed grid resolution, which is consistent with the Kelvin-Helmholtz instability property. 
They also compared the energy spectrum for the turbulent kinetic energy and magnetic field energy (see Fig.~\ref{fig:chapter16_Aguilera-Miret2020b}). 
They concluded that the sub-grid model with $(C_M,C_N,C_T)=(8,0,0)$ can capture more efficient magnetic field amplification than the direct simulation counterpart until $5~{\rm ms}$ after the merger. This sub-grid setup is also able to mimic the magnetic-field energy amplification 
of the direct simulation with $\Delta x=74~{\rm m}$ at $10~{\rm ms}$ after the merger. Also, they reported how sensitive the remnant massive neutron star evolution is to the choice of the pre-factor. If they choose $(C_M,C_N,C_T)=(8,1,1)$, $(8,2,2)$, $(8,4,4)$, or $(8,8,8)$, the Kelvin-Helmholtz amplification is less efficient than $(C_M,C_N,C_T)=(8,0,0)$. More importantly, it triggers the collapse of a black hole, which is not observed in the direct simulations. They concluded that the choice of the non-zero value of $C_N$ and $C_T$, particularly the latter, facilitates the angular momentum transport too efficiently, which results in a spuriously early black hole formation. 

Besides the subtleness of the choice of the pre-factor, there are a couple of caveats: (i) it is not clear which term(s) in $H^i_M$ in Eq.~(\ref{eq:chpater16_SGS}) leads the efficient amplification during the Kelvin-Helmholtz instability phase. (ii) The magnetic field energy shown in Fig.~\ref{fig:chapter16_Aguilera-Miret2020a} may not be consistent with their pre-merger magnetic field since it should be $\sim 10^{40}~{\rm erg}~B_{11.7}^2 R_6^3$ where $B_{11.7}=B/10^{11.7}{\rm G}\approx B/(5\times 10^{11}{\rm G})$ and $R_6=R/10^6{\rm cm}$. (iii) The scale of the energy spectrum in Fig.~\ref{fig:chapter16_Aguilera-Miret2020b} may not be consistent with their grid setup. The highest wavenumber of $k\approx 10^{-3}~{\rm cm^{-1}}$ is the spatial scale of $60~{\rm m}$, but their employed grid resolution is $\Delta x=147~{\rm m}$. 

Ruiz and his collaborators reported new magnetized binary neutron star merger simulations~\cite{Ruiz:2021qmm}. The pre-merger magnetic field is the pulsar-like dipole field and confined dipole field with presumably $\approx 10^{16.5}$--$10^{17}~{\rm G}$. The employed grid resolution is $\Delta x=90~{\rm m}$--$110~{\rm m}$. The neutron star is modeled with the SLy equation of sate~\cite{Douchin:2001sv} and H4 equation of state~\cite{Glendenning:1991es}. 
They explored the prompt black hole formation case and delayed collapse case. For the latter, the collapse time is $\approx 10$--$40~{\rm ms}$ after the merger. Their findings are qualitatively the same as their series works~\cite{Ruiz:2016rai,Ruiz:2017inq,Ruiz:2017due,Ruiz:2019ezy,Ruiz:2020via}. An incipient jet was not found for the prompt collapse case, irrespective of the model. For the delayed collapse case, the magnetic field energy is amplified to $\sim 10^{51}~{\rm erg}$ during the Kelvin-Helmholtz instability phase. In some models, it is furthermore amplified up to $\sim 10^{52}~{\rm erg}$ until the black hole formation. The amplification highly depends on the magnetic field topology. Except for one model (H4M2.8I in their terminology), which is the same as in Ref.~\cite{Kiuchi:2014hja} other than the pre-merger magnetic field strength and grid setup, they found a launching incipient jet. Besides the caveats mentioned above, one notable difference from the simulations in Ref.~\cite{Aguilera-Miret:2020dhz}, in which a much weaker pre-merger magnetic field is assumed, is the saturation energy of the magnetic field $\sim 10^{50}~{\rm erg}$ at the end of the Kelvin-Helmholtz instability phase. 

Palenzuela and his collaborators reported new magnetized binary neutron star simulations with the gradient sub-grid scale model~\cite{Palenzuela:2021gdo}. The pre-merger magnetic field is $5\times 10^{11}~{\rm G}$. They performed an in-depth resolution study with $\Delta x=30~{\rm m}$, $60~{\rm m}$, and $120~{\rm m}$ with and without the sub-grid model. The neutron star is modeled with the APR equation of state~\cite{Akmal:1998cf}. 
The pre-factor is chosen $(C_M,C_N,C_T)=(8,0,0)$. Their findings are summarized as follows: (i) With the sub-grid model, the highest and middle-resolution runs indicate the convergence to $\bar{B}_{\rm tor}\approx 4$--$5\times 10^{15}~{\rm G}$ and $\bar{B}_{\rm pol}\approx 5$--$6\times 10^{15}~{\rm G}$ at the end of the Kelvin-Helmholtz instability phase in the core region defined by $\rho \ge 10^{13}~{\rm g~cm^{-3}}$. 
It is likely to be consistent with the result in the highest direct simulation. Such a trend is also observed as $\bar{B}_{\rm tor}\approx \bar{B}_{\rm pol} \approx 4$--$5\times 10^{14}~{\rm G}$ in the envelope region defined by $10^{10}~{\rm g~cm^{-3}} \le \rho \le 10^{13}~{\rm g~cm^{-3}}$. (ii) The spectrum analysis suggests the kinetic turbulent energy spectrum follows the Kolmogorov power law with $\propto k^{-5/3}$ and the magnetic energy spectrum follows the Kazantsev power law with $\propto k^{3/2}$. The inverse cascade is likely to occur, which results in the increase of the coherent length scale of the magnetic field from $\approx 0.7~{\rm km}$ to $2~{\rm km}$ at the end of the simulation of $50~{\rm ms}$ after the merger. 
(iii) During the first $50~{\rm ms}$ at least, the efficient angular momentum redistribution is not likely to be facilitated, implying the MRI is not operating because there is no static, large-scale background field over which we can define an unstable perturbation of the MRI. 

However, there are a couple of caveats: (i) They employed a cut-off density of $6\times 10^{13}~{\rm g~cm^{-3}}$ below which the sub-grid term turns off. This cut-off density seems to be significant, and the effect seems not to be negligible for the evolution of both the remnant core and envelope. (ii) Since the highest direct simulation is likely to agree with the converged sub-grid simulation, it implies that the direct simulation looks like entering the convergent regime. It should be confirmed in a higher-resolution direct simulation. Otherwise, we cannot conclude that the saturation magnetic field energy is physical or a consequence of the sub-grid prescription. (iii) The role of the MRI should be explored in a more extended time-scale simulation because their simulation suggests the onset of the magnetic winding. Particularly, several diagnostics to explore the MRI in the non-linear phase proposed in Ref.~\cite{Hawley:2011tq} should be investigated (see also Ref.~\cite{Kiuchi:2017zzg}). (iv) Since the high resolution of $\Delta x=60~{\rm m}$ still needed for the convergence in the simulation with the sub-grid model, there might be a ``double" counting for the turbulence, i.e., one developed by the direct simulation and the other by the sub-grid model which mimics the direct simulation. It is necessary to quantify how such an artifact would affect the convergent property.

Aguilera-Miret and his collaborators reported new magnetized binary neutron star merger simulations with the gradient sub-grid scale model~\cite{Aguilera-Miret:2021fre}. The pre-merger magnetic field is $10^{12}~{\rm G}$ with an aligned dipole, misaligned dipole, and multipole topology. They also explored a strong magnetic field case with $10^{15}~{\rm G}$. The employed grid resolution is $\Delta x =60~{\rm m}$. The neutron star is modeled by the APR equation of state~\cite{Akmal:1998cf}. The pre-factor is chosen $(C_M,C_N,C_T)=(8,0,0)$. Their findings are summarized as follows: (i) At the end of the Kelvin-Helmholtz instability, the averaged magnetic field strength in the core region converges within a factor of three for the toroidal component and two for the poloidal component irrespective of the pre-merger field topology and field strength. Consequently, the magnetic field energy saturates within a factor of three, and the strong pre-merger magnetic field results in a similar saturation level. (ii) The evolution of the energy spectrum is insensitive to the pre-merger magnetic field topology. The coherent length of the magnetic field evolves from $\sim 0.8~{\rm km}$ to $2~{\rm km}$ at the end of the simulation of $30~{\rm ms}$ after the merger. Therefore, the pre-merger field topology memory is likely to be lost during the merger, implying the universality of their result. 

A caveat is that they only consider the purely poloidal magnetic field, implying zero magnetic helicity. However, the magnetic field configuration after the merger remnant settles down to the quasi-equilibrium state could depend on the net magnetic field helicity because the magnetic helicity conserved in the ideal MHD framework links to the field topology~\cite{Moffatt:1978}.

Sun and his collaborators reported new magnetized binary neutron star merger simulations with neutrino physics~\cite{Sun:2022vri}. The pre-merger magnetic field is a dipole-like field with presumably $\approx 10^{16.5}$--$10^{17}~{\rm G}$ (see Eq.~(\ref{eq:chapter16_dipole})). The employed grid resolution is $\Delta x=110~{\rm m}$. The neutron star is modeled with the SLy equation of states~\cite{Douchin:2001sv}. During the simulation, they employed a piecewise polytrope equation of state for the cold part combined with the analytic expression of the thermal part. They also employ the M1 scheme for the neutrino radiation field. 
Their finding is qualitatively similar to what they did in the past~\cite{Ruiz:2016rai,Ruiz:2017inq,Ruiz:2017due,Ruiz:2019ezy,Ruiz:2020via,Ruiz:2021qmm}: an incipient jet launching for the delayed collapse case irrespective of the neutrino effect. 
They estimated the neutrino viscous and drag effect on the MRI for the first time in the magnetized binary neutron star merger simulation and reported the effect is not significant. They also reported that neutrino cooling mitigates the baryon-loading in the funnel region above the black hole as reported in Ref.~\cite{Mosta:2020hlh}. The angular momentum loss due to the neutrino emission is minor during their simulation time, $\approx 15~{\rm ms}$ after the black hole formation. 

The caveat, except for those in their series works, is that the luminosity for the heavy-lepton neutrino suddenly increases up to $\sim 10^{53}~{\rm erg~s^{-1}}$ at $\approx 5~{\rm ms}$ after the black hole formation, which seems not to be consistent with the other neutrino radiation transfer simulation of binary neutron star mergers~\cite{Kiuchi:2022nin,Sekiguchi:2015dma,Sekiguchi:2016bjd}. 

Palenzuela and his collaborators reported magnetized binary neutron star merger simulations with neutrino physics and the sub-grid model~\cite{Palenzuela:2022kqk}. They implemented a finite temperature nuclear theory-based equation of states HS~\cite{Shen:1998gq} and LS220~\cite{Lattimer:1991nc} and neutrino leakage scheme~\cite{OConnor:2009iuz}. The pre-merger magnetic field is $10^{11}~{\rm G}$. They employed a grid resolution of $\Delta x=187~{\rm m}$. 
The pre-factor in the sub-grid model is set to be $(C_M,C_N,C_T)=(0.5,0,0)$. They concluded that although they found an efficient amplification of the magnetic field from $10^{11}~{\rm G}$ to $\sim 10^{14}~{\rm G}$ for $\sim 8~{\rm ms}$ after the merger in the sub-grid model simulation the magnetic field hardly alters the neutrino emission such as the neutrino luminosity. 

The caveat is that their result on the neutrino emission is not likely conclusive because (i) the previous work showed that at least the grid resolution of $\Delta x=60~{\rm m}$ is necessary to obtain the saturated magnetic field of $\sim 10^{16}~{\rm G}$ in the sub-grid model run with $(C_M,C_N,C_T)=(8,0,0)$~\cite{Aguilera-Miret:2021fre} and (ii) the simulation timescale is not long enough to explore the MHD effect on the neutrino radiation. 

Kiuchi and his collaborators reported a new implementation for advanced HLLD Riemann solver~\cite{Mignone:2008ii} and applied it to a magnetized binary neutron star merger with neutrino physics~\cite{Kiuchi:2022ubj}. In the context of the accretion disk, it has been well known the HLL(E) Riemann solver~\footnote{It is known that HLL(E) and Local-Lax-Friedrich (LLF) Riemann solver~\cite{Amiram:2006zjz,Kurganov:2000ovy} give the essentially same result.} commonly used in the numerical relativity community is very diffusive~\cite{Mignone:2020qec,White:2015omx}. They explored how the less diffusive HLLD Riemann solver affects the post-merger evolution. They embedded a large-scale poloidal magnetic field of $10^{15}~{\rm G}$ inside a massive torus formed after a short-lived remnant massive neutron star collapses into a black hole. The employed grid resolution is $\Delta x=150~{\rm m}$. The neutron star is modeled with the SFHo equation of state~\cite{Steiner:2012rk}. 
Their findings are summarized as follows: (i) The small-scale magneto-turbulence due to the MRI is not able to be sustained if we employ the HLLE solver because of inherently large numerical diffusivity. (ii) As a result, the large-scale magnetic field of $\sim 10~{\rm km}$ is artificially enhanced in the simulation with the HLLE Riemann solver compared to that with the HLLD Riemann solver (see Fig.~26 in Ref.~\cite{Kiuchi:2022ubj}). (iii) More energetic (but spurious) Poynting flux-dominated outflow is launched in the simulation with the HLLE Riemann solver compared to that with the HLLD Riemann solver (see Fig.~27 in Ref.~\cite{Kiuchi:2022ubj}). 

The caveat is that they only implemented the third-order piecewise parabolic method for the reconstruction in the Riemann problem~\cite{Colella:1984}. Therefore, they did not quantify how the HLLE Riemann solver with a higher-order reconstruction such as the MP5~\cite{Suresh:1997} would reduce the numerical diffusion in binary neutron star merger simulations. 

Combi and Siegel reported a new magnetized binary neutron star merger simulations with neutrino physics~\cite{Combi:2022nhg}. They employed finite temperature nuclear equation of states, LS220, SFHo, and APR~\cite{Akmal:1998cf,Lattimer:1991nc,Steiner:2012rk} and a zeroth-momentum (M0) scheme based on a ray-by-ray for the neutrino radiation transport~\cite{Radice:2016dwd}. The pre-merger magnetic field is $5\times 10^{15}~{\rm G}$. The employed grid resolution is $\Delta x=180~{\rm m}$ and $220~{\rm m}$. They reported that the magnetic field does not significantly impact the dynamical ejecta. They also reported that the MRI is fully developed inside the remnant massive neutron star and facilitates the angular momentum transport. 

The caveat is as follows: (i) Many previous works reported the efficient Kelvin-Helmholtz instability and fully resolving MRI are challenging with their setup~\cite{Ciolfi:2017uak,Ciolfi:2019fie,Kiuchi:2017zzg,Kiuchi:2023obe,Mosta:2020hlh,Palenzuela:2021gdo}. (ii) Because of the shortness of their simulation timescale of $\approx 10~{\rm ms}$, the MHD effect, particularly the role of the angular momentum transport inside the remnant, needs to be clarified. Note that the MRI-driven viscous timescale or the magnetic winding timescale is longer than their simulation time. 
Also, the detailed analysis on the MRI is necessary. 

De Haas and his collaborators reported magnetized binary neutron star merger simulations with neutrino physics~\cite{deHaas:2022ytm}, a follow-up work of Ref.~\cite{Mosta:2020hlh}. They added a dipole-like large-scale magnetic field to a remnant massive neutron star:
\begin{align}
A_r=A_\theta=0,~A_\phi=B_0\sin\theta\frac{r_{\rm falloff}^3}{r_{\rm falloff}^3+r^3},
\end{align}
and they varied $B_0=10^{13}~{\rm G}$, $10^{14}~{\rm G}$, $10^{15}~{\rm G}$, $5\times10^{15}~{\rm G}$ and $r_{\rm falloff}=5~{\rm km}$, $10~{\rm km}$, $15~{\rm km}$, $20~{\rm km}$. The employed grid resolution is $\Delta x=185~{\rm m}$. The neutron star is modeled with the LS220 equation of state~\cite{Lattimer:1991nc}. They employed the leakage scheme for the neutrino cooling~\cite{Mosta:2020hlh} and parameterized prescription for the neutrino heating. They found in the two simulations with $(B_0,r_{\rm falloff})=(10^{15}~{\rm G},20~{\rm km})$ and $(5\times 10^{15}~{\rm G},10~{\rm km})$ the jet is launched and the velocity distribution of the ejecta has a fast component with $0.4$--$0.6~{\rm c}$. 

The caveat is that (i) the role of the MRI, particularly low magnetic field cases, needs to be clarified more, and (ii), as they mentioned, the large-scale field inside the merger remnant is an assumption.

\begin{figure*}[t]
         \includegraphics[width=0.99\linewidth]{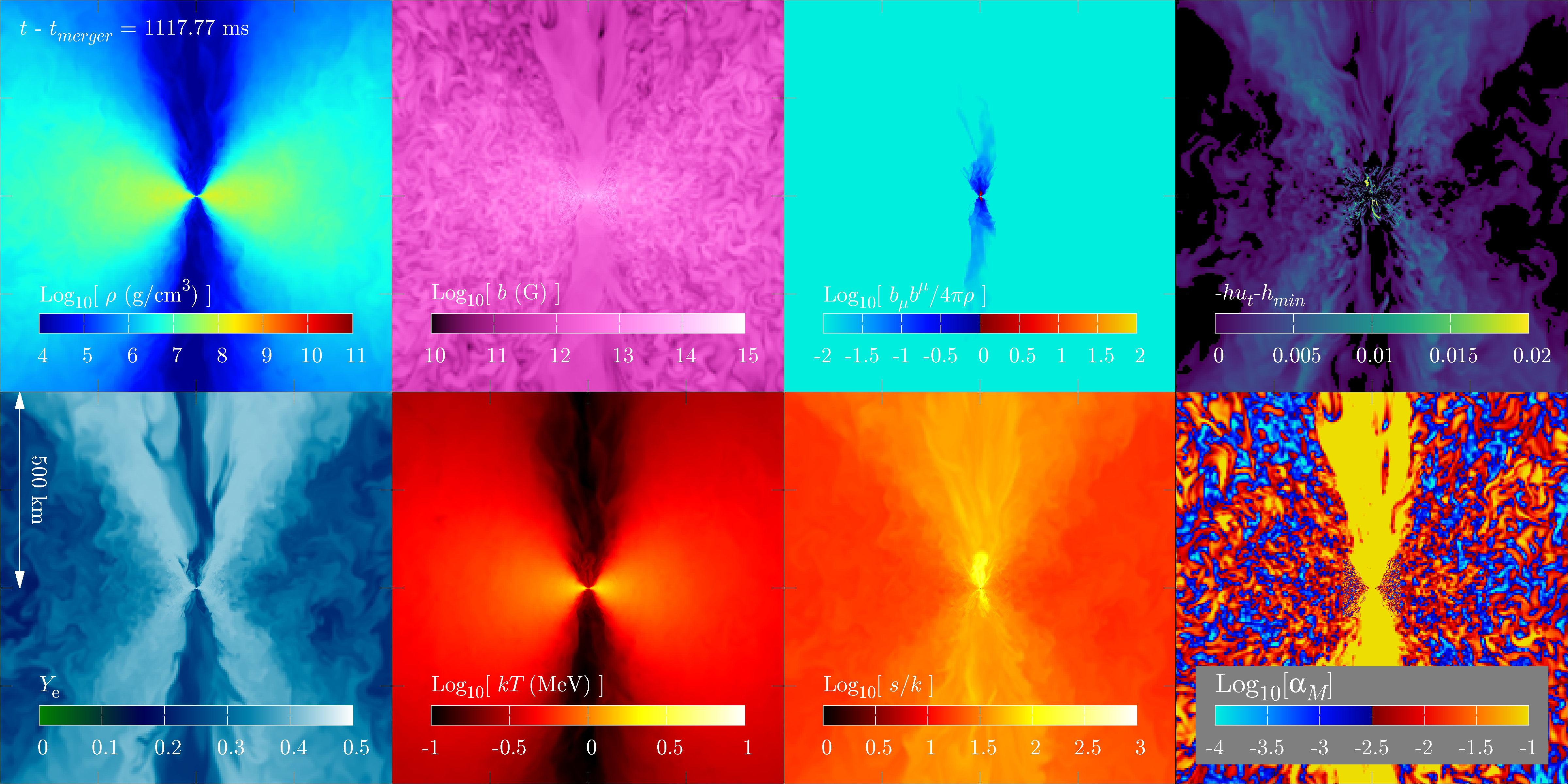}
         \caption{Profiles for rest-mass density (top-left), magnetic-field strength (top-second from left), magnetization parameter (top-second from right), unboundedness defined by the Bernoulli criterion (top-right), electron fraction (bottom-lef\
t), temperature (bottom-second from left), entropy per baryon (bottom-second from right), and Shakura-Sunyaev $\alpha_\mathrm{M}$ parameter (bottom-right) on the $y-y_\mathrm{AH}=0$ plane at $t-t_\mathrm{merger}\approx 1.1$~s. See also the movie: \url{http://www2.yukawa.kyoto-u.ac.jp/~kenta.kiuchi/anime/FUGAKU/out_SFHo_12_15.mp4}. The figure is taken from Ref.~\cite{Kiuchi:2022nin}.
      }\label{fig:chapter16_Kiuchi2023a}
\end{figure*}

\begin{figure*}
         \includegraphics[scale=0.23]{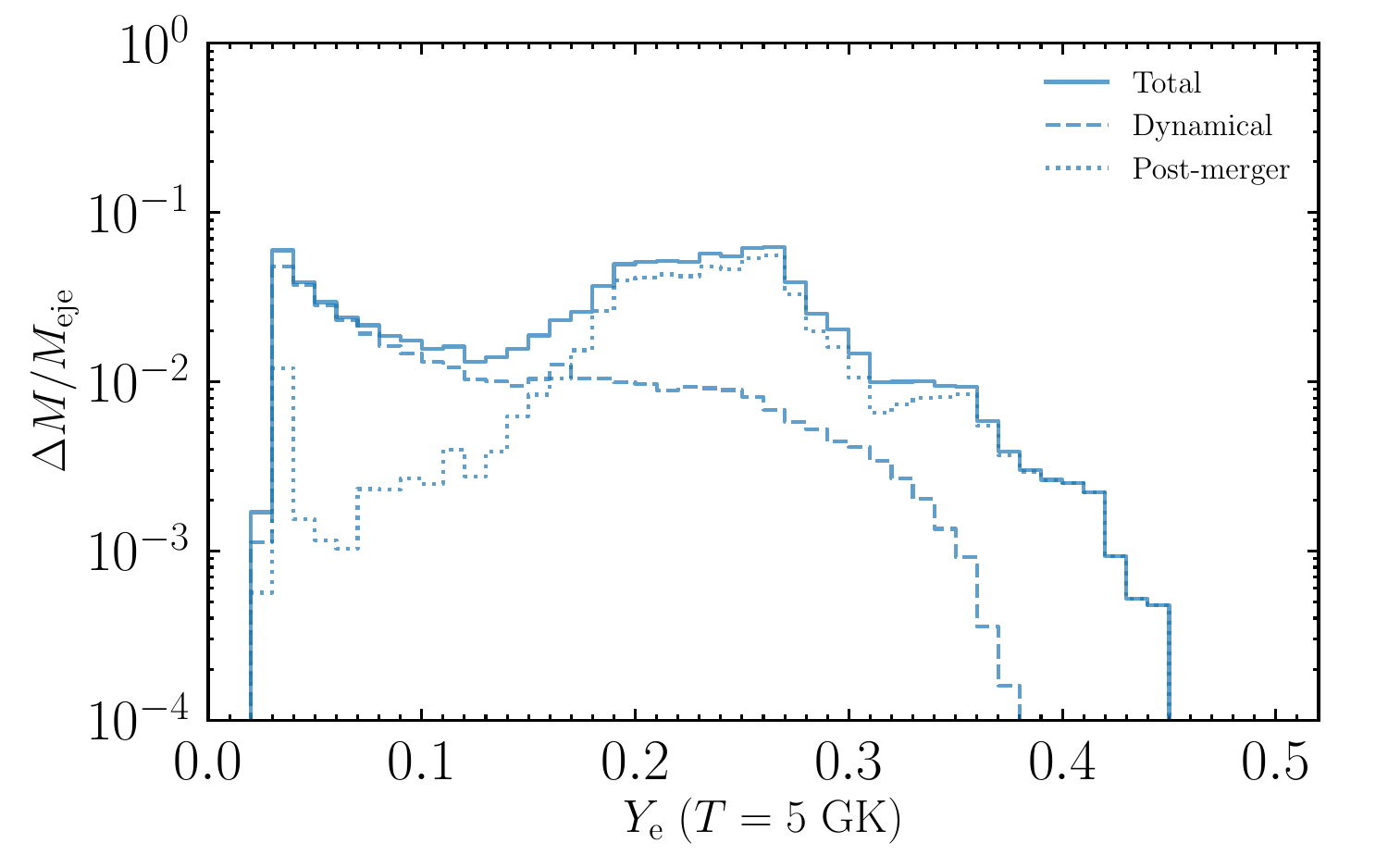}
         \includegraphics[scale=0.23]{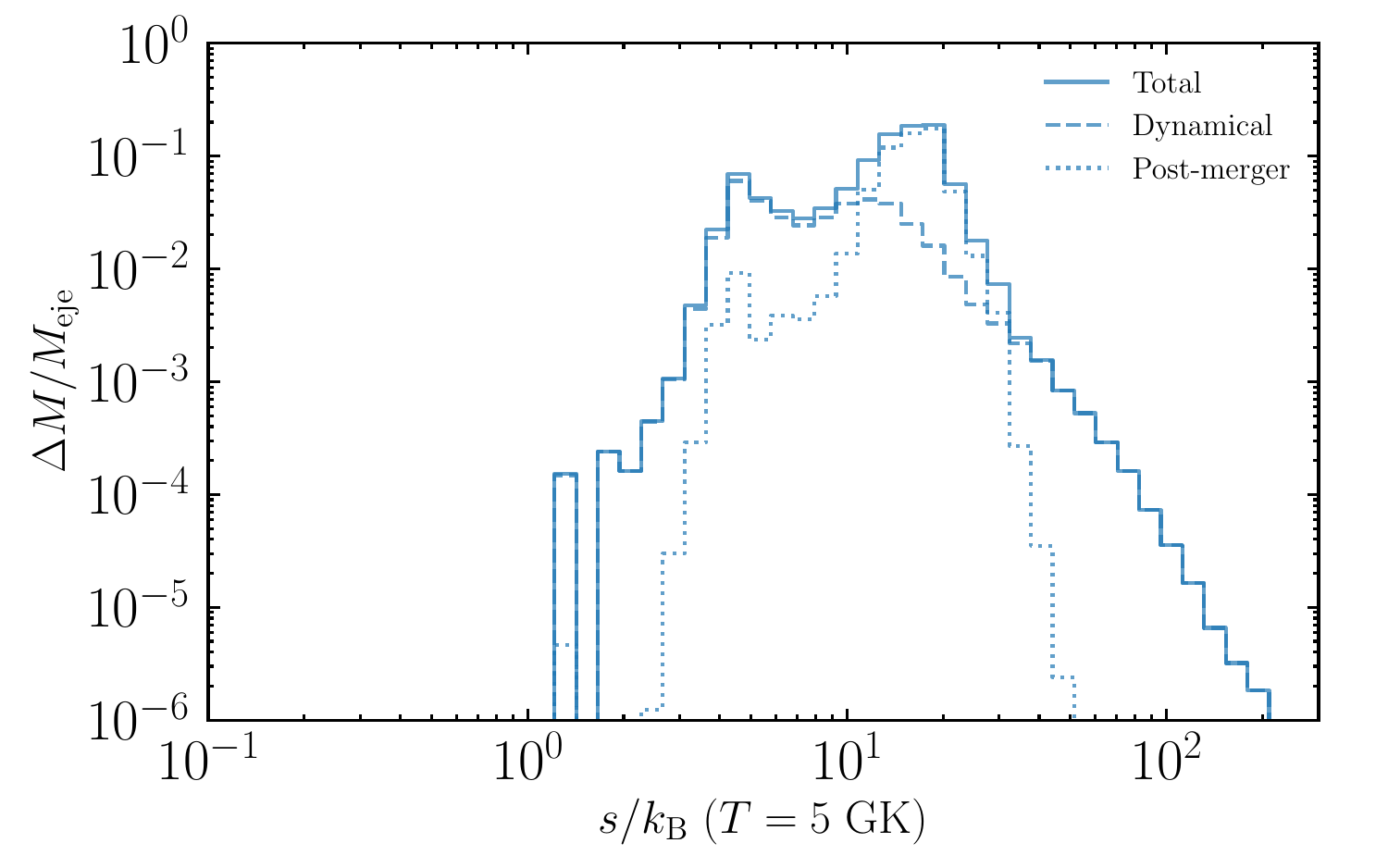}
         \includegraphics[scale=0.23]{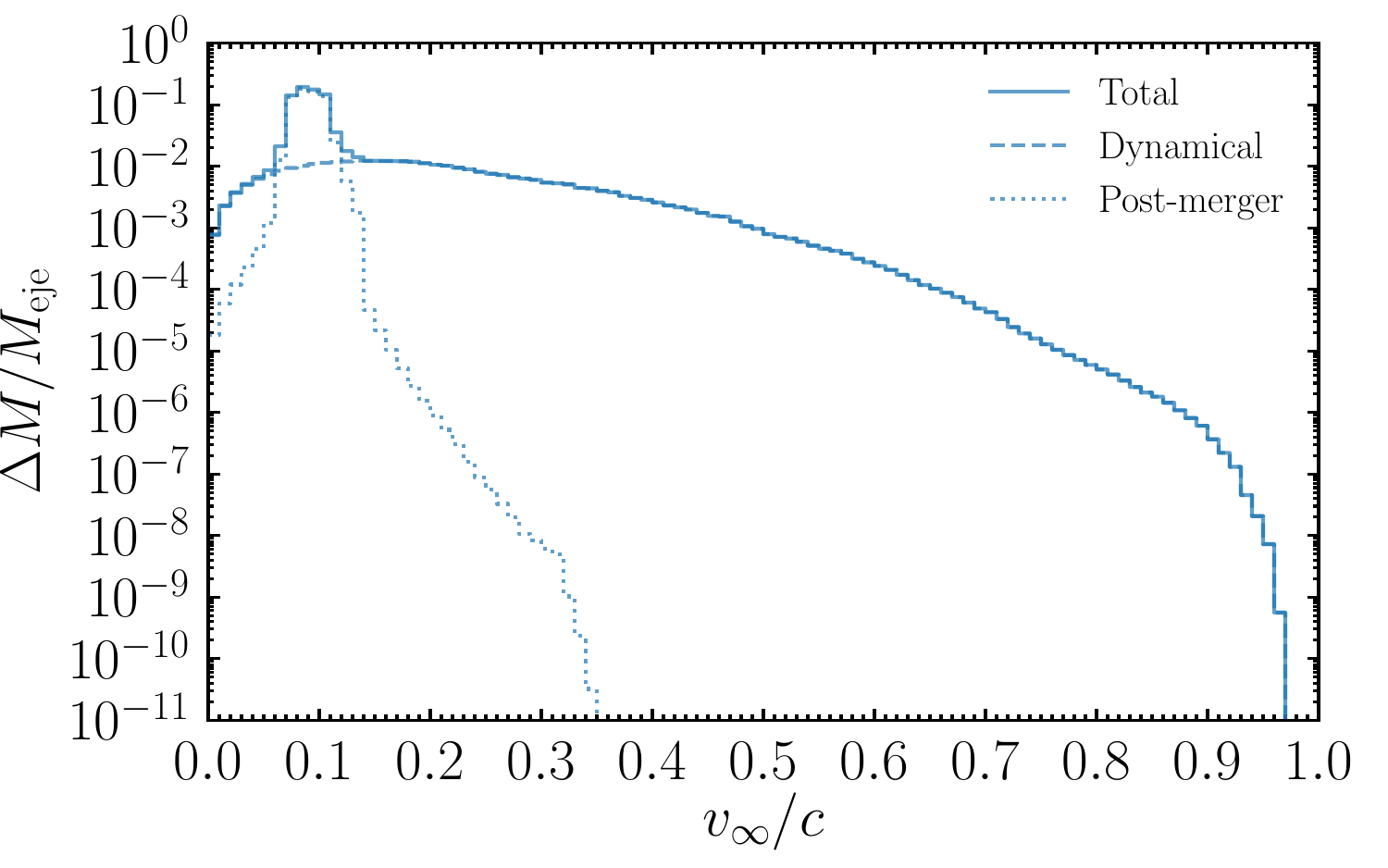}
         \caption{
         Mass histogram of the ejecta as a function of the electron fraction (top-left), the entropy per baryon (top-right), and the terminal velocity (bottom) at $t-t_\mathrm{merger}\approx 1.1~{\rm s}$ calculated by the tracer particle. The solid, dashed, and dotted curves denote the profiles for the total, dynamical, and post-merger ejecta, respectively. The figure is taken from Ref.~\cite{Kiuchi:2022nin}.
         }\label{fig:chapter16_Kiuchi2023b}
\end{figure*}

Kiuchi and his collaborators reported new magnetized binary neutron star merger simulations with neutrino physics~\cite{Kiuchi:2022nin}. The pre-merger magnetic field is $10^{15}~{\rm G}$. The employed grid resolution is $\Delta x=150~{\rm m}$ and $200~{\rm m}$ for the convergence test. The neutron star is modeled with the SFHo equation of state~\cite{Steiner:2012rk}. They performed a simulation up to one second after the merger (see Fig.~\ref{fig:chapter16_Kiuchi2023a} for the final snapshot on a meridional slice). Their findings are summarized as follows: (i) After the short-lived massive neutron star collapses into a black hole, the magnetic field inside the massive torus is amplified by the axisymmetric MRI and magnetic winding. As a result, at $\sim 0.1~{\rm s}$ after the merger, the fully turbulent state due to the MRI is established, and the turbulent viscosity with the Shakura-Sunyaev parameter of $0.01$--$0.03$ facilitates the angular momentum transport. 
The turbulent state is sustained by the MRI dynamo, proved by the butterfly diagram lasting until the end of the simulation (see Fig.~3 in Ref.~\cite{Kiuchi:2022nin}). 
(ii) Due to the angular momentum transport and turbulent viscous heating, the torus expands, and the temperature drops. 
As a result, the neutrino cooling becomes inefficient. (iii) Finally, the post-merger mass ejection due to the MRI-driven turbulent viscosity sets in, and the mass of $\approx 8\times 10^{-3}M_\odot$ is ejected from the torus (see Fig.~\ref{fig:chapter16_Kiuchi2023b} for the detailed properties of the ejecta). (iv) The jet launching is not observed until the end of the simulation of $\approx 1~{\rm s}$ after the merger. (v) The convergence in terms of the grid resolution is almost achieved, implying the simulation quality could be good enough to compare to the observational data such as AT 2017gfo. 

The caveat is summarized as follows: (i) Since with their setup, it is impossible to resolve an efficient Kelvin-Helmholtz instability and the axisymmetric MRI inside the short-lived remnant massive neutron star whose lifetime is $\approx 17~{\rm ms}$, there is a possibility that the large-scale magnetic field is build up before the black hole formation. (ii) Although the fallback motion lasts and resultant ram pressure overcomes the magnetic pressure until the end of the simulation of $\approx 1~{\rm s}$, the fallback time scale could depend on the binary neutron star model because the other works employing a different binary neutron star model reported the fallback motion ceases at $O(0.01)~{\rm s}$ after the black hole formation (see Ref.~\cite{Ruiz:2021qmm} for example). (iii) The absence of the jet launching could be caused by spurious spin down of a black hole due to an insufficient grid resolution. The resolution study with the Cowling approximation suggested it is unlikely, though. 

Chabanov and his collaborators explored pre-merger magnetic field topology in magnetized binary neutron star mergers~\cite{Chabanov:2022twz}. The pre-merger magnetic field is a traditional confined dipole field with $\sim 10^{14}~{\rm G}$ and ``crustal" field with $\sim 2\times 10^{14}~{\rm G}$. The employed grid resolution is $\Delta x=70~{\rm m}$ and $105~{\rm m}$ for the convergence test. The neutron star is modeled with the TNTYST equation of state~\cite{Togashi:2017mjp}. Their finding is summarized as follows: (i) The amplification process comprises the Kelvin-Helmholtz instability phase, the subsequent decay phase, and the subsequent turbulent phase. (ii) At the end of the Kelvin-Helmholtz instability phase, the magnetic field energy for the ``crustal" configuration is smaller by a factor of few than the confined configuration. At the end of the turbulent phase, the former is smaller by a one-order magnitude than the latter. They conclude that the ``crustal" configuration leads to inefficient Kelvin-Helmholtz instability compared to the confined configuration widely used in magnetized binary neutron star mergers. 

The caveat is that the saturation energy of the magnetic field for the ``crustal" configuration needs to be more extensively explored since the high-resolution simulation shows a significant growth rate, implying the physical saturation could be far from what they reported. 

\begin{figure*}
    \centering
    \includegraphics[width=0.95\textwidth]{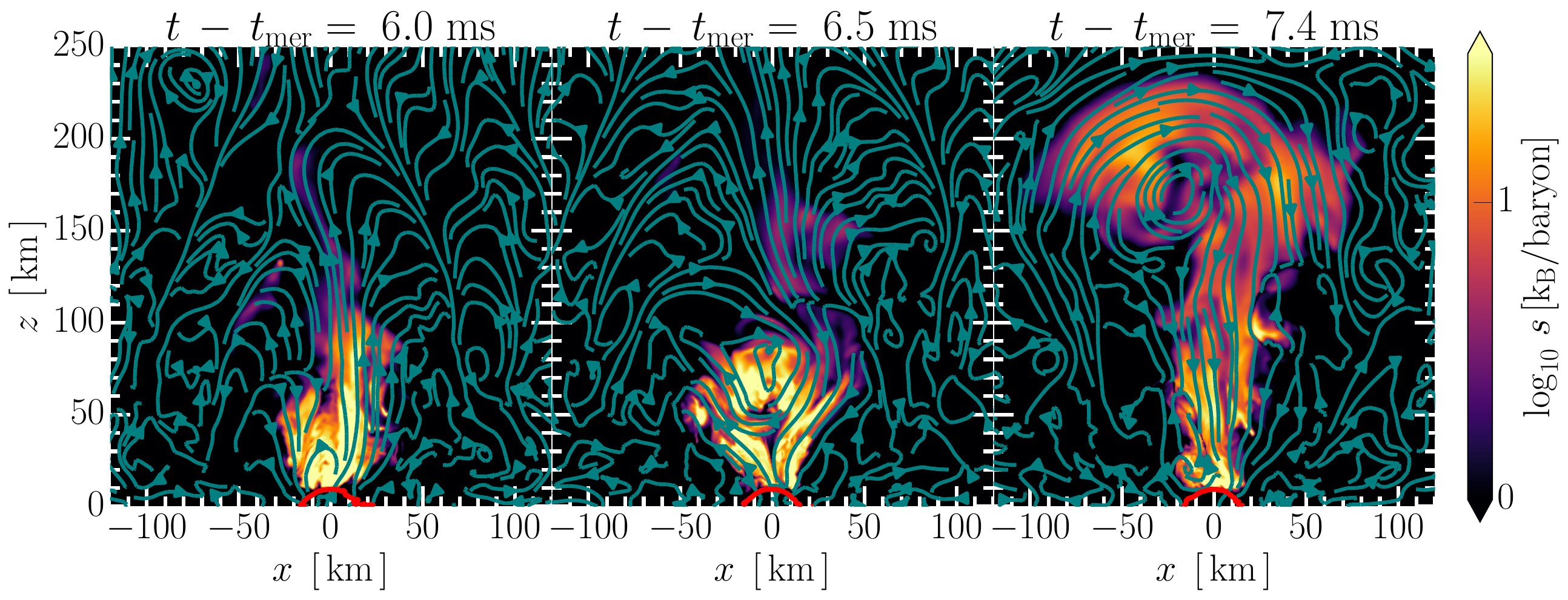}
    \caption{Flaring process at the surface (red) of the neutron star merger remnant. The entropy per baryon $s$ is shown in color (indicating heating inside the flare) with the magnetic field lines. The time is defined as the post-merger time. 
    {(Left)} Built-up of magnetic stresses due to buoyancy of field
    lines from the star. {(Center)} Inflation of the connected flux tube
    due to strong differential rotation at the surface of the star, where
    both footpoints are anchored. {(Right)} Detachment of
  the flare within $\simeq 1 \rm ms$ of its creation. In the simulations, this cycle repeats quasi-periodically until a steady outflow is launched. The model is $\sigma=0.01$. 
  The figure is taken from Ref.~\cite{Most:2023sft}.}
    \label{fig:chapter16_Most2023a}
\end{figure*}

\begin{figure}
  \centering
  \includegraphics[width=0.65\textwidth]{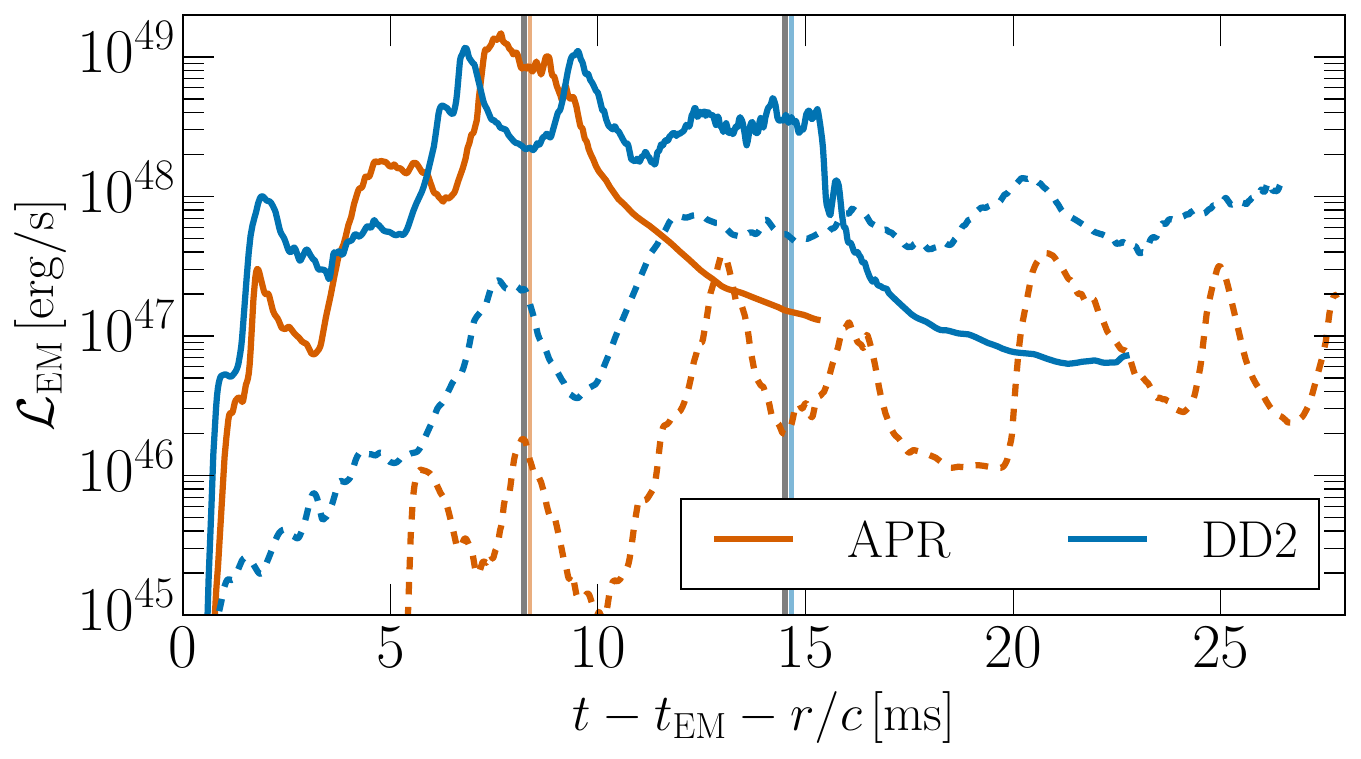}
  \caption{Electromagnetic luminosity as a function of the retarded time with $t_{\rm EM}$, which corresponds to the time of the first electromagnetic outburst extracted at a radius $r=236~\rm km$ from the stellar remnant. Solid lines refer to high magnetization models, $\sigma=0.01$, and dashed lines to low magnetization models, $\sigma = 0.001$. For the high magnetization cases, black hole formation is indicated by a vertical black line. The figure is taken from Ref.~\cite{Most:2023sft}.}
  \label{fig:chapter16_Most2023b}
\end{figure}

Most and Quataert reported new magnetized binary neutron star merger simulations with neutrino physics and a sub-grid model to reproduce a large-scale dynamo~\cite{Most:2023sft}. Specifically, their sub-grid model is $\alpha$ dynamo described by
\begin{align}
e^\mu = \kappa b^\mu,
\end{align}
where $e^\mu$ and $b^\mu$ are the electric and magnetic fields in a fluid comoving frame. In the non-relativistic ideal MHD approximation, $e^i=E^i+({\bf V}\times {\bf B})^i=0$ (see Eqs.~(\ref{eq:chapter16_induction})--(\ref{eq:chapter16_alphaomega})). $\kappa$ is a calibration parameter such that the saturation value of the magnetization parameter $\sigma=b^2/\rho$ is controlled. They performed two cases with $\sigma\approx 0.01$ and $\approx 0.001$. The pre-merger magnetic field is $10^{15}~{\rm G}$. The employed grid resolution is $\Delta x =250~{\rm m}$. The neutron star is modeled by the DD2 and APR equation of sates~\cite{Akmal:1998cf,Hempel:2009mc}. The leakage scheme incorporates neutrino cooling. 
Figure~\ref{fig:chapter16_Most2023a} summarizes their result. Due to the $\alpha$ dynamo prescription, the strong magnetic field of $\sim 10^{17}~{\rm G}$ is produced, and the resultant magnetic buoyancy force pushes the fluid elements upward near the surface (left panel). A strongly magnetized loop that sticks out of the stellar surface is formed. It is twisted by a differential rotation of the merger remnant and inflate (center panel). As the twist is increased, the inflated bubble detaches from the merger remnant (right panel). Since this process repeats, a periodicity is imprinted in the Poynting flux as shown in Fig.~\ref{fig:chapter16_Most2023b}. The luminosity and the periodicity strongly depend on the equation of state and the dynamo calibration parameter $\kappa$. 

The caveat is as follows: (i) More detailed calibration $\kappa$ or detailed modeling of the dynamo prescription is necessary because the result is susceptible to it. (ii) The role of the MRI needs to be more clarified. 

Combi and Siegel reported new magnetized binary neutron star merger simulations with neutrino physics~\cite{Combi:2023yav}. The pre-merger magnetic field is $3\times 10^{15}~{\rm G}$. The employed grid resolution is $\Delta x=180~{\rm m}$. The neutron star is modeled with the APR equation of state~\cite{Akmal:1998cf}. The neutrino treatment is the same as the previous work~\cite{Combi:2022nhg}. Their finding is summarized as follows: (i) After the Kelvin-Helmholtz phase, the inverse turbulent cascade creates a large-scale magnetic field. (ii) The large-scale toroidal magnetic field is further amplified, and an incipient jet is launched. Also, the post-merger mass ejection due to the MHD effect sets in. 

The caveat is as follows: (i) Since many previous simulations revealed that an efficient Kelvin-Helmholtz instability and the MRI are hard to be resolved by their setup, the mechanism to generate the large-scale magnetic field must be clarified more. (ii) Because of the lack of a resolution study, it is hard to quantify systematic errors in their findings, such as electromagnetic field luminosity and post-merger mass ejection, due to the grid resolution. This point is crucial to compare the simulation result with the observation data. 

\begin{figure}[t]
\centering
\includegraphics[width=0.6\linewidth]{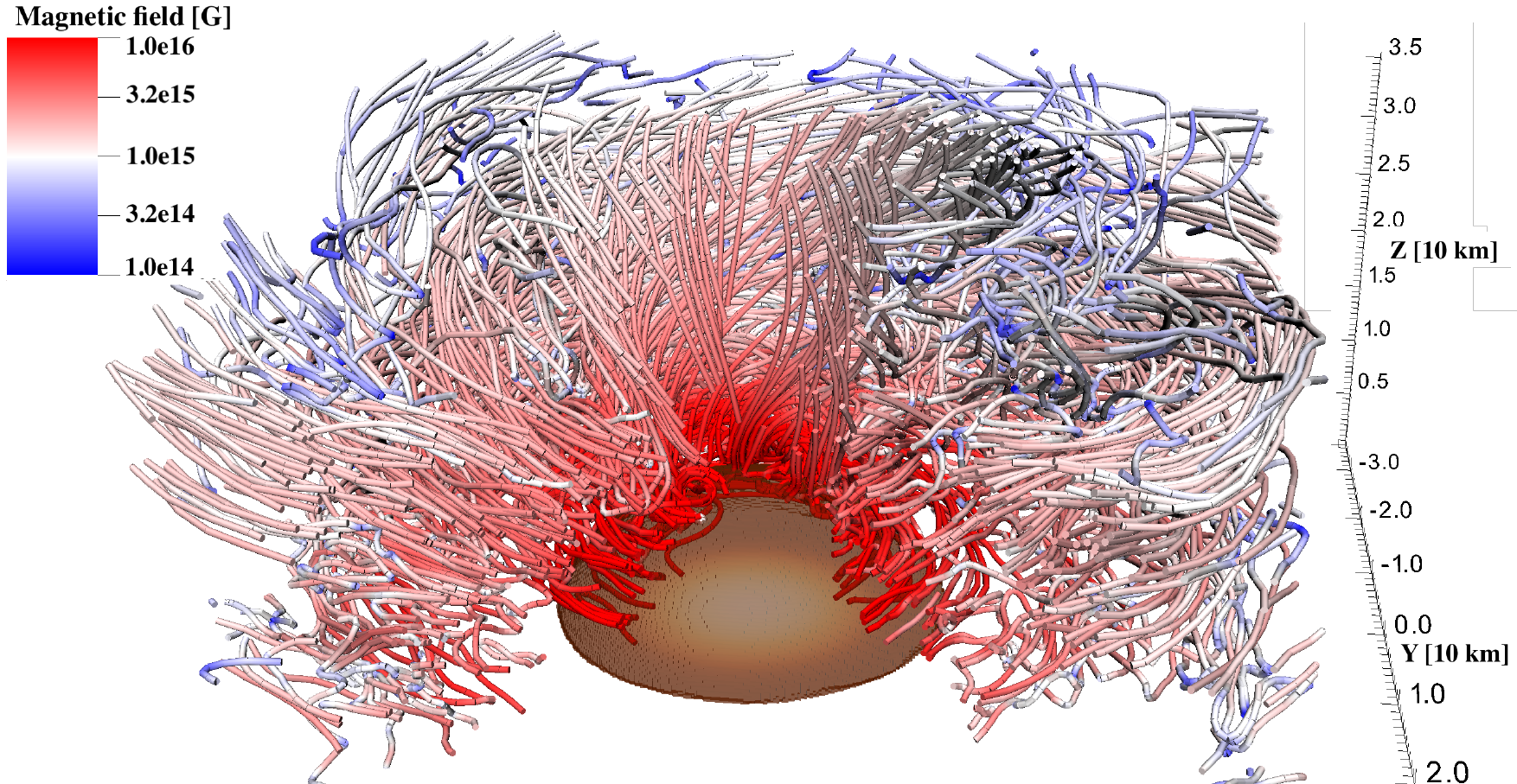}
\caption{Magnetic field lines for the density of $\rho < 10^{13}$\,g\,cm$^{-3}$ at $t-t_\mathrm{merger}\approx 130$~ms. The core of the hypermassive neutron star is shown for the density of $\rho > 10^{13}$\,g\,cm$^{-3}$. The figure is taken from Ref.~\cite{Kiuchi:2023obe}. 
}\label{fig:chapter16_Kiuchi2024a}
\end{figure}

\begin{figure}[h!]%
\centering
\includegraphics[width=0.49\textwidth]{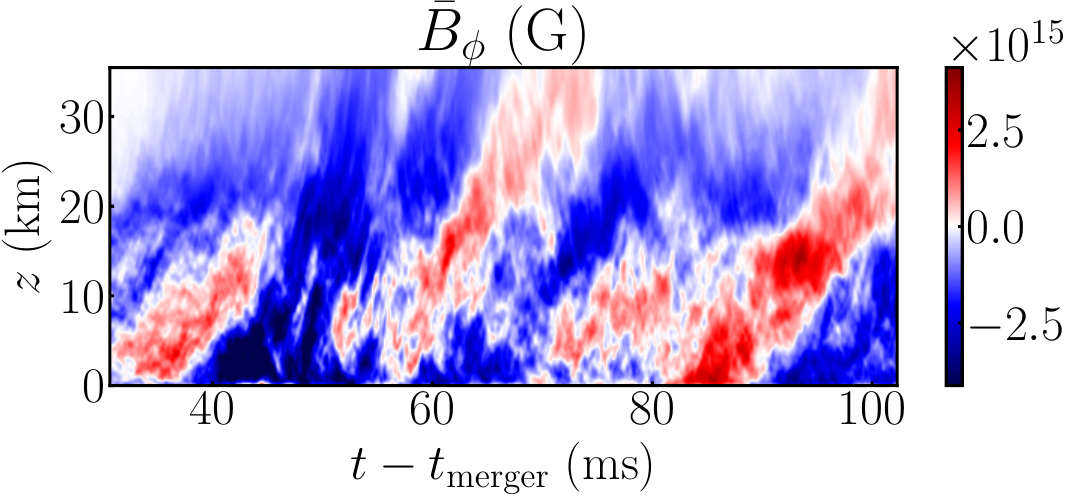}
\includegraphics[width=0.49\textwidth]{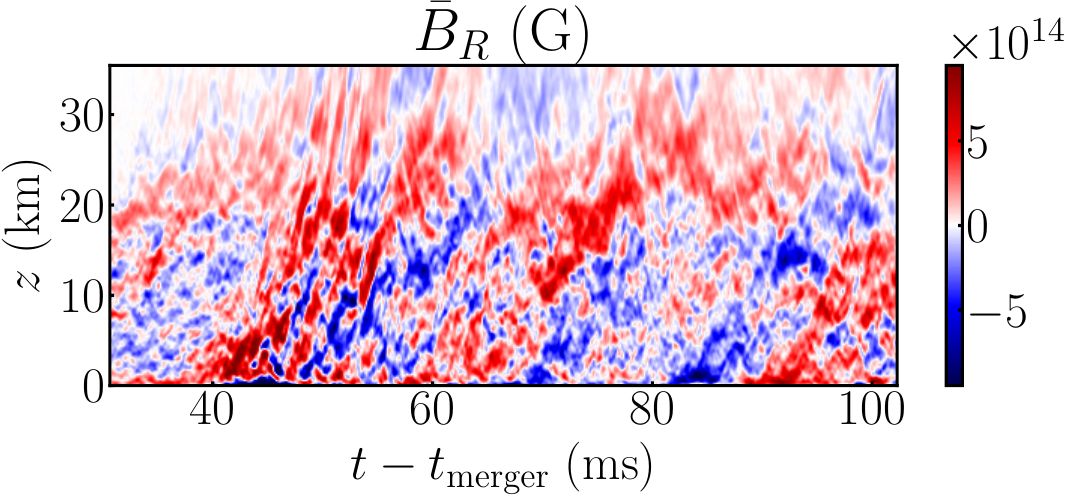}\\
\includegraphics[width=0.49\textwidth]{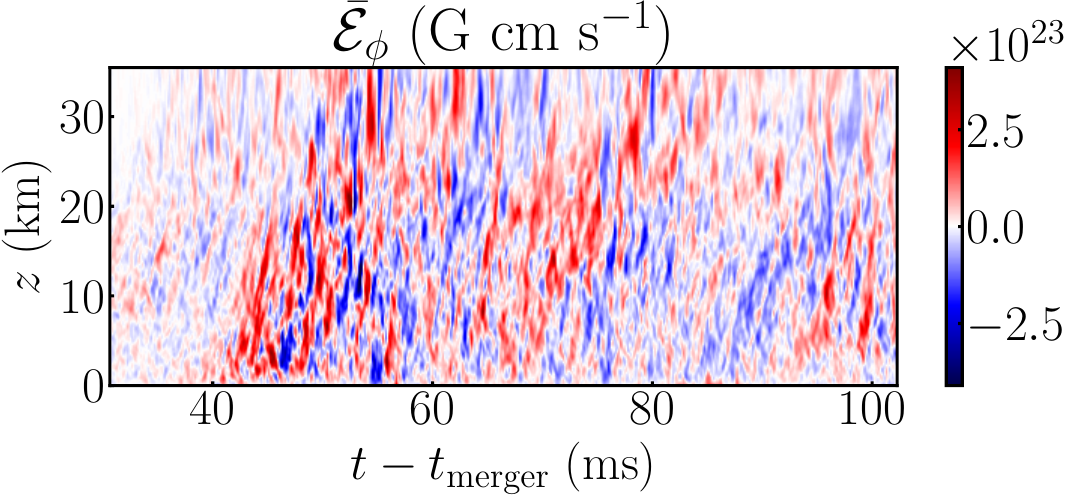}
\includegraphics[width=0.49\textwidth]{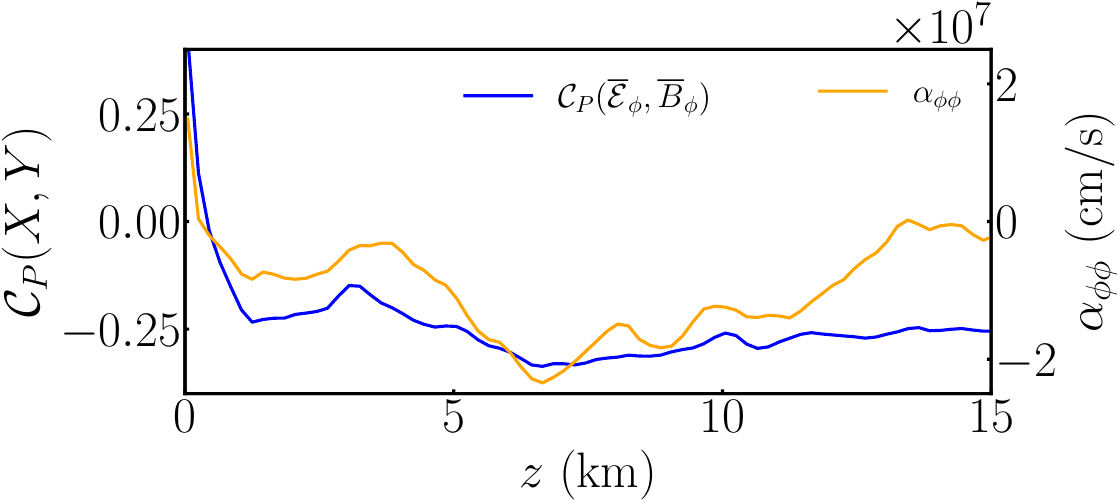}
\caption{Butterfly diagram at $R = 30$ km: (Top-left) Mean toroidal magnetic field $\bar{B}_\phi$. (Top-right) Mean radial magnetic field $\bar{B}_R$. (Bottom-left) Toroidal electromotive force $\bar{\mathcal{E}}_\phi$. (Bottom-right) $\alpha_{\phi\phi}$ parameters (orange) and correlation between $\bar{\mathcal{E}}_\phi$ and $\bar{B}_\phi$ (blue). The figure is taken from Ref.~\cite{Kiuchi:2023obe}. 
}\label{fig:chapter16_Kiuchi2024b}
\end{figure}

\begin{figure}[h!]%
\centering
\includegraphics[width=0.64\textwidth]{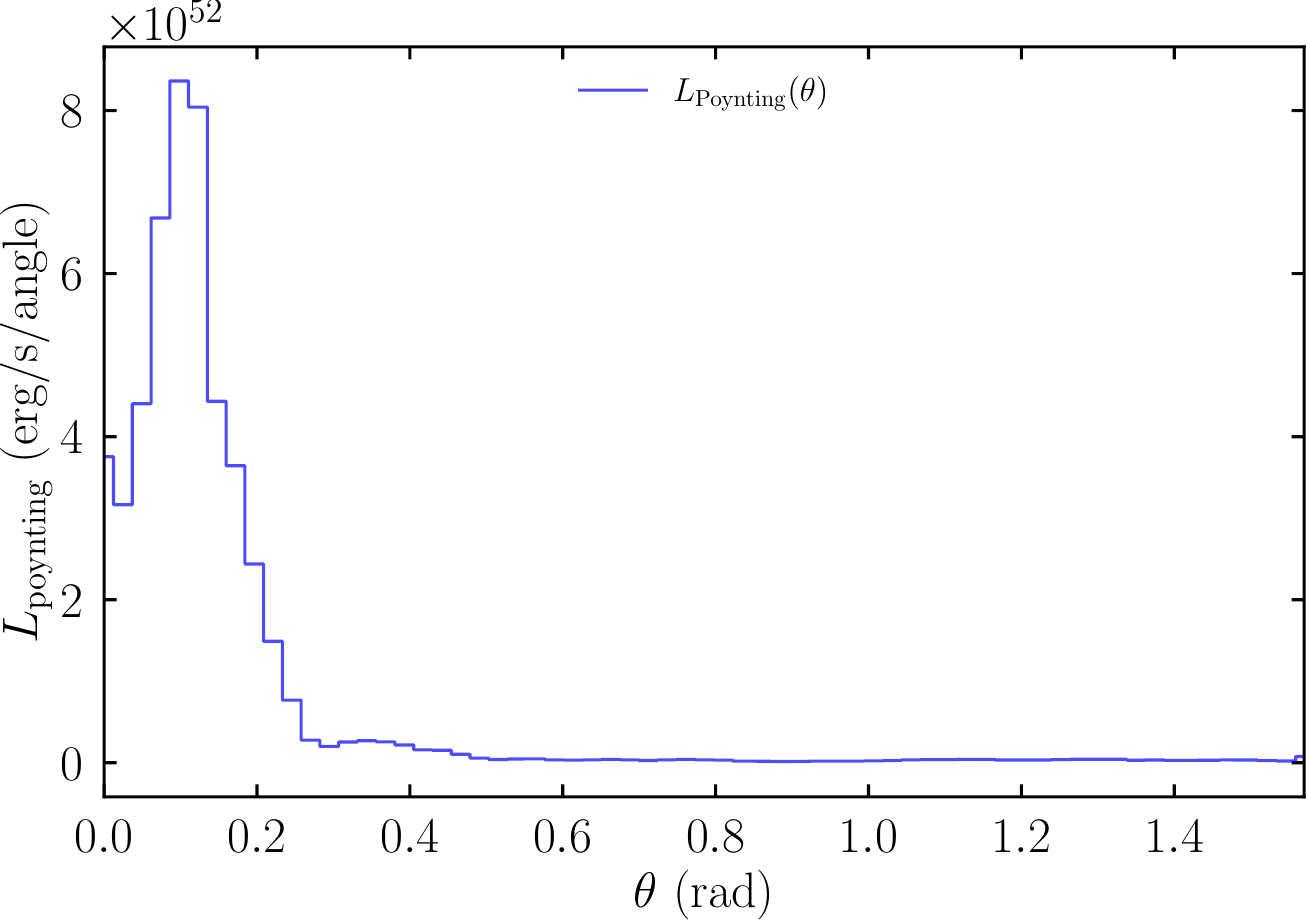}\\
\includegraphics[width=0.64\textwidth]{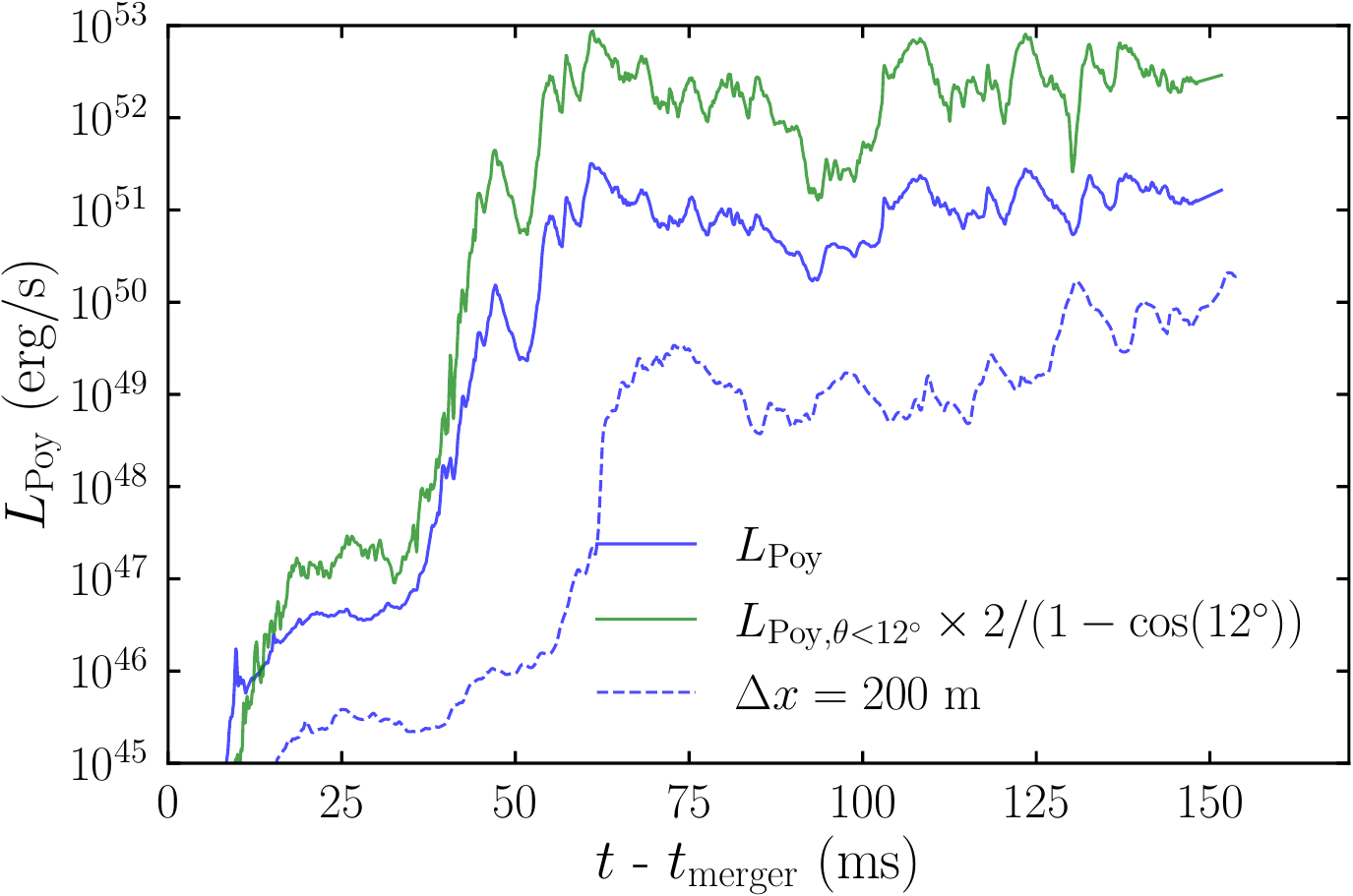}\\
\includegraphics[width=0.64\textwidth]{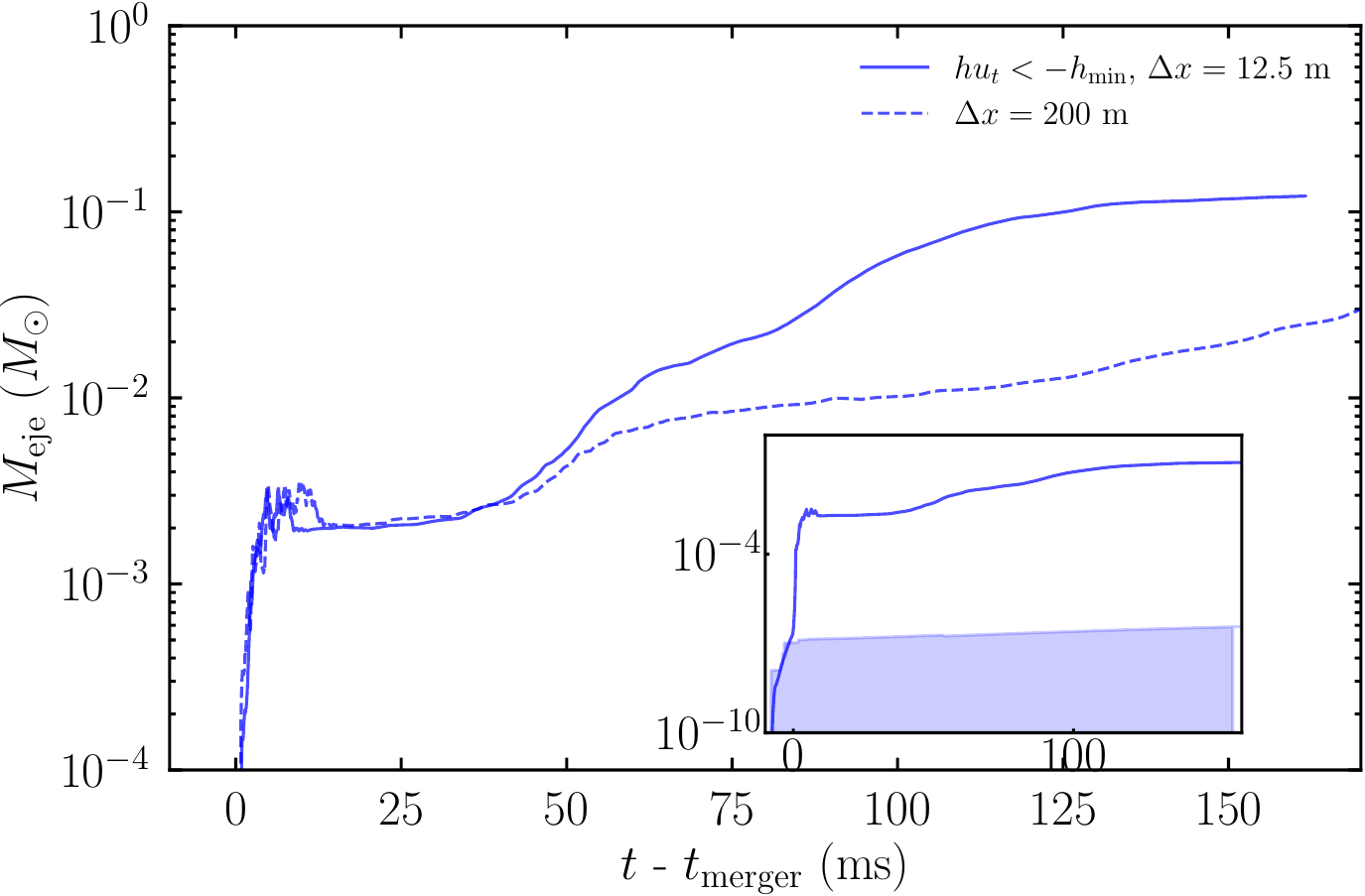}
\caption{(Top) Angular distribution of the luminosity of the Poyinting flux at the end of the simulation of $t-t_\mathrm{merger}\approx 150$~ms. 
(Middle) Luminosity for the Poynting flux as a function of the post-merger time. The green curve is the jet-opening-angle corrected luminosity. The blue-dashed curve plots the luminosity for the simulation with $\Delta x_{\rm finest}=200$~m. 
(Bottom) Ejecta as a function of the post-merger time. The solid curve denotes the ejecta satisfying the Bernoulli criterion. The colored region in the inset shows the violation of the baryon mass conservation.
The blue-dashed curve plots the ejecta for the simulation with $\Delta x_{\rm finest}=200$~m. The figure is taken from Ref.~\cite{Kiuchi:2023obe}. 
}\label{fig:chapter16_Kiuchi2024c}
\end{figure}

Kiuchi and his collaborators reported new magnetized binary neutron star merger simulations with neutrino physics~\cite{Kiuchi:2023obe}. 
The pre-merger magnetic field is $3\times 10^{15}~{\rm G}$. The employed grid resolution is $\Delta x=12.5~{\rm m}$ from the inspiral to the first $\approx 30~{\rm ms}$ after the merger, subsequently $\Delta x =50~{\rm m}$ up to $\approx 50~{\rm ms}$ after the merger, and finally $\Delta x=100~{\rm m}$ until the end of the simulation of $175~{\rm ms}$ after the merger (see Method section in Ref.~\cite{Kiuchi:2023obe} for detail). 
They also performed the convergence test with $\Delta x=200~{\rm m}$. 
The neutron star is modeled with the DD2+Timmes equation of state~\cite{Hempel:2009mc,Timmes:2000}, which results in the long-lived remnant massive neutron star formation. The neutrino radiation transport is solved with the M1+GR Leakage scheme. 
Their findings are summarized as follows: (i) As reported previously~\cite{Chabanov:2022twz,Kiuchi:2014hja,Kiuchi:2015sga,Kiuchi:2017zzg}, the low resolution with $\Delta x=200~{\rm m}$ is unable to capture the efficient Kelvin-Helmholtz instability and the MRI (see Extended Data Figs.~1 and 2 in Ref.~\cite{Kiuchi:2023obe}). The high-resolution simulation with $\Delta x=12.5~{\rm m}$ can sustain the MRI-driven turbulence. 
(ii) The neutrino viscosity and drag are likely to be irrelevant in binary neutron star mergers because of the efficient Kelvin-Helmholtz instability (see Extended Data Fig.~3 in Ref.~\cite{Kiuchi:2023obe}, which solved the dispersion relations~(\ref{eq:chaper16_dis1})--(\ref{eq:chaper16_dis2}) on top of the simulation data). (iii) Because the MRI-driven turbulence is responsible for generating the electromotive force~(\ref{eq:chapter16_induction}) and the period in the butterfly diagram agrees with the prediction by the $\alpha\Omega$ dynamo theory,
\begin{align}
P_{\rm theory}=2\pi\left|\frac{1}{2}\alpha_{\phi\phi}\frac{d\Omega}{d\ln R}k_z\right|^{-1/2},
\end{align}
where $k_z$ is the wavenumber corresponding to the pressure scale height (see also Eq.~(\ref{eq:chapter16_alphaomega}) for $\alpha_{\phi\phi}$), 
the $\alpha\Omega$ dynamo can be interpreted as a mechanism for the large-scale magnetic field generation as shown in Figs.~\ref{fig:chapter16_Kiuchi2024a}--\ref{fig:chapter16_Kiuchi2024b}\footnote{The working hypothesis to derive Eqs.~(\ref{eq:chapter16_alphaomega1})--(\ref{eq:chapter16_alphaomega3}) is directly verified in Ref.~\cite{Kiuchi:2023obe}}. 
(iv) The pre-merger large-scale magnetic field is harmless for the large-scale dynamo because such a relic magnetic field stays buried deep inside the merger remnant core throughout the simulation, and the dynamo wave appears from the surface of the remnant core (see Extended Data Fig.~5 in Ref.~\cite{Kiuchi:2023obe}). 
(v) The relativistic Poynting-flux dominated outflow with the luminosity of $\sim 10^{51}~{\rm erg~s^{-1}}$ is launched by the large-scale magnetic field due to the $\alpha\Omega$ dynamo. Also, the Lorentz force due to the large-scale field drives an enormous post-merger mass ejection of $\approx 0.1M_\odot$ as shown in Fig.~\ref{fig:chapter16_Kiuchi2024c}. It should be noted that the low-resolution simulation with $\Delta x=200~{\rm m}$ underestimates the luminosity of the Poynting flux by two orders of magnitude and the post-merger ejecta mass by one order of magnitude at $\approx 100~{\rm ms}$ after the merger, which corresponds to a ``longest"-term simulation among the previous simulations conducted to date for the long-lived case. Therefore, the systematic error due to the grid resolution is astonishingly large, which is crucial for the comparison to the observational data such as AT 2017gfo. 

The caveat is that (i) it is necessary to confirm this picture by a simulation starting from a much weaker pre-merger magnetic field since, as claimed in Refs.~\cite{Aguilera-Miret:2023qih,Palenzuela:2021gdo}, the MRI may not be effective at least early post-merger phase because of the absence of the static and large-scale background field. (ii) The other dynamo such as the Taylor-Spruit dynamo~\cite{Spruit:2001tz} could be effective in generating the large-scale magnetic field, particularly deep inside the merger remnant core because such a region is not subject to the MRI due to the positive radial gradient of the angular velocity. (iii) Since the magnetic Prandtl number determined by the numerical viscosity and resistivity is an order of unity in their simulation, the large-scale dynamo property may change in the high magnetic Prandtl number regime. 

Aguilera-Miret and his collaborators reported new magnetized binary neutron star merger simulations with the gradient sub-grid scale model~\cite{Aguilera-Miret:2023qih}. The pre-merger magnetic field strength is $5\times 10^{11}~{\rm G}$. They also performed a follow-up simulation with the ``crustal" configuration proposed in Ref.~\cite{Chabanov:2022twz}. 
The employed grid resolution is $\Delta x=60~{\rm m}$. The neutron star is modeled by the APR equation of state~\cite{Akmal:1998cf}. 
The remnant massive neutron star is long-lived. The pre-factor of the sub-grid model is $(C_M,C_N,C_T)=(8,0,0)$ with the cut-off density of $2\times 10^{11}~{\rm g~cm^{-3}}$ below which the sub-grid terms are turned off. 
As consistent with their previous works~\cite{Aguilera-Miret:2021fre,Palenzuela:2021gdo}, the Kelvin-Helmotholz instability triggers the turbulent magnetic field amplification up to $\sim 10^{50}~{\rm erg}$ until $\sim 5~{\rm ms}$ after the merger. The energy spectrum analysis suggests that isotropic turbulence results in comparable strength in the poloidal and toroidal magnetic fields. After the Kelvin-Helmholtz instability phase, the turbulent resistivity is enhanced. As a result, the small-scale magnetic field is diffused, and the magnetic field with a coherent length of a few ${\rm km}$ is developed. Because of the resultant coherent poloidal magnetic field, the magnetic winding works as a further amplification of the toroidal field, and the magnetic field energy ends up at $\sim 10^{51}~{\rm erg}$ at $\sim 110~{\rm ms}$ after the merger. Although they observed a helicoidal structure of the magnetic field, they did not find a jet launching until the end of the simulation of $\approx 110~{\rm ms}$ after the merger. The MRI potentially unstable region inside the remnant massive neutron star has a highly non-axisymmetric intensity, implying the prediction of $\lambda_{\rm MRI}$ is a non-trivial task because the classical and widespread way of evaluation of $\lambda_{\rm MRI}$ assumes a background and homogeneous field. 
Even starting with the ``crustal" configuration, they found a similar saturation energy of $\sim 10^{50}~{\rm erg}$ at the end of the Kelvin-Helmholtz instability phase, which may imply the result in Ref.~\cite{Chabanov:2022twz} could be merely caused by an insufficient resolution. 

The caveat is that (i) there is still room for the MRI investigation because they did not explicitly show that the MRI does not emerge in their simulation, although the MRI diagnostics, such as the MRI quality factor or the Maxwell stress, indicate the emergence of the MRI. 
(ii) The role of the turbulent resistivity needs to be more clarified because of the inherently large numerical resistivity of their Local-Lax-Freidrich Riemann solver (see Figs.~12 and 13 in Ref.~\cite{Kiuchi:2022ubj} for the magnetic reconnection problem with different Riemann solvers). (iii) The $\alpha\Omega$ dynamo's role needs to be clarified, particularly a correlation between the electromotive force and mean field as demonstrated in Ref.~\cite{Kiuchi:2023obe}. 

\begin{figure}
    \centering
    \includegraphics[width=0.65\textwidth]{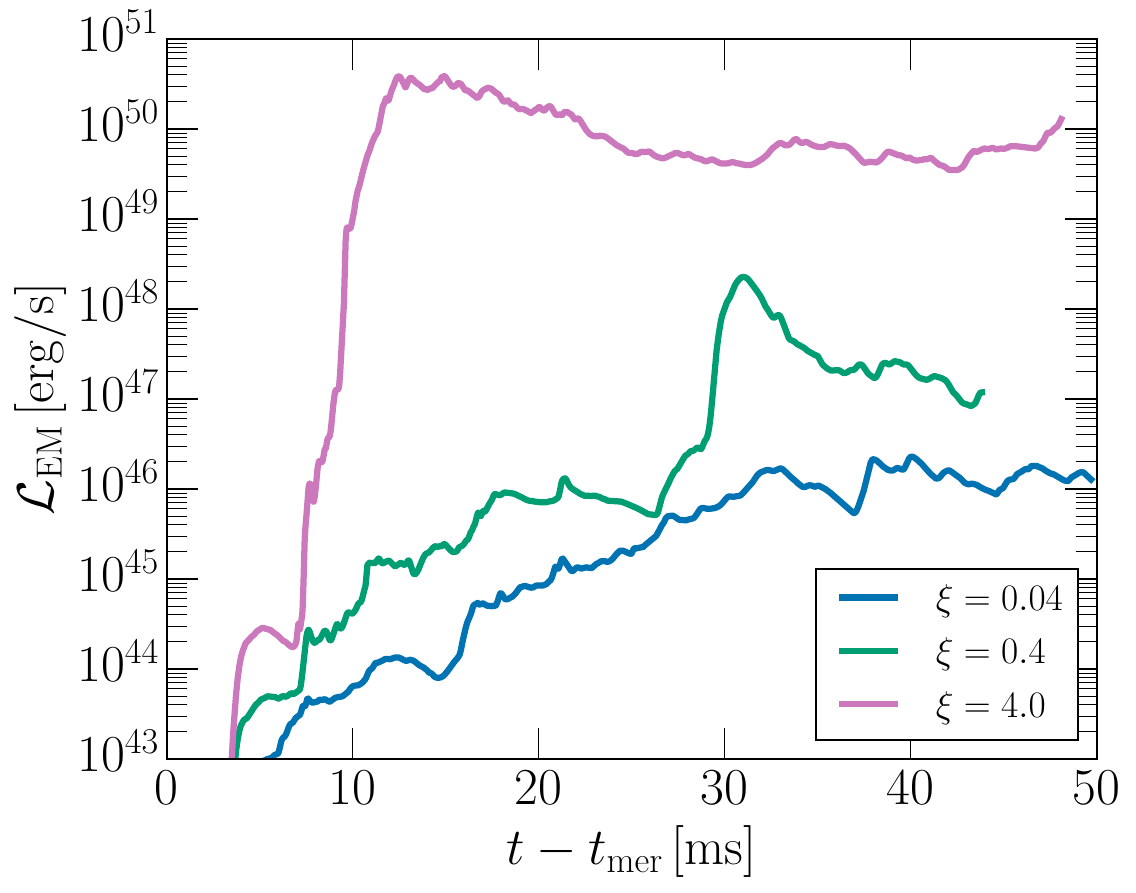}
    \caption{Electromagnetic (Poynting) flux, $\mathcal{L}_{\rm EM}$,
  measured at $r= 236\, \rm km$ from the remnant, as a function of the post-merger time. Different curves
correspond to the different dynamo parameters $\xi$. The figure is taken from Ref.~\cite{Most:2023sme}. }
    \label{fig:chapter16_Most2024}
\end{figure}

Most proposed a new sub-grid model for the $\alpha\Omega$ dynamo in magnetized binary neutron star merger context~\cite{Most:2023sme}. By assuming the dynamo effects grow relative to the resistive timescale in the Ohm's law, he arrived at a tensorial relation between the electric field $e^\mu$ and magnetic field $b^\mu$ in the fluid co-moving frame:
\begin{align}
e^\mu = {\kappa^\mu}_\nu b^\nu,
\end{align}
where ${\kappa^\mu}_\nu=-\eta {\alpha^\mu}_\nu$ with the resistivity $\eta$ and dynamo alpha (see also Eq.~(\ref{eq:chapter16_alphaomega})).  
This equation is furthermore simplified by assuming that ${\kappa^\mu}_\nu=\kappa({\delta^\mu}_\nu+u^\mu u_\nu)$ where $u^\mu$ is a fluid four velocity:
\begin{align}
e^\mu=\kappa b^\mu. 
\end{align}
The dynamo coefficient $\kappa$ and the dynamo saturation are inspired by high-resolution magnetized binary neutron star merger simulations~\cite{Kiuchi:2017zzg,Kiuchi:2023obe}:
\begin{align}
&\kappa=\kappa_{\rm HMNS} \max(0,\Delta_{\rm turb}),\\
&\Delta_{\rm turb} = 1 - \frac{\sigma}{\sigma_{\rm turb}},\\
&\sigma_{\rm turb} = \xi \left(l^{\rm HMNS}_{\rm MRI}\right)^2 \left(\frac{\Delta x}{12.5~{\rm m}}\right)^2 \sigma^{\mu\nu} \sigma_{\mu\nu},\\
&l^{\rm HMNS}_{\rm MRI} = \max(0,a \log_{10}(\rho/\rho_*) \exp[-|b\log_{10}(\rho/\rho_*)|^{5/2}])~{\rm m},
\end{align}
where $\kappa_{\rm HMNS} \approx 0.025-0.035$ is the dynamo parameter inferred from the ultra-high resolution simulation~\cite{Kiuchi:2023obe}, $\sigma$ is the magnetization parameter, $\xi$ is a parameter, $\sigma_{\mu\nu}$ is the shear tensor, and $l^{\rm HMNS}_{\rm MRI}$ is the MRI wavelength inside the remnant massive neutron star. 
An important assumption here is that the $\alpha\Omega$ dynamo will terminate once a fraction $\xi$ of the turbulent kinetic energy is converted into the magnetic field energy. 
$l^{\rm HMNS}_{\rm MRI}$ is fitted by the global simulation in Ref.~\cite{Kiuchi:2017zzg} with $a=22.31984$, $b=-0.425832$, and $\rho_*=1.966769\times 10^9~{\rm g~cm^{-3}}$~\cite{Radice:2020ids}. He left $\xi$ as a free parameter and performed a simulation with $\xi=0.04$, $0.4$, and $4$~\footnote{In principle, $\xi$ should be smaller than unity.}. The pre-merger magnetic field is $10^{15}~{\rm G}$. The employed grid resolution is $\Delta x=200~{\rm m}$. The result is qualitatively similar to his previous work~\cite{Most:2023sft} (see Fig.~\ref{fig:chapter16_Most2023a} and its explanation). Figure~\ref{fig:chapter16_Most2024} shows how the choice of $\xi$ results in the luminosity for the Poynting flux, and it indicates $\xi=4$ is closed to those found in Ref.~\cite{Kiuchi:2023obe} (see Fig.~\ref{fig:chapter16_Kiuchi2024c}). 

The caveat is that a further calibration for the sub-grid model for the $\alpha\Omega$ dynamo is necessary because the result is sensitive to the choice of $\xi$.

\section{Summary and prospect}\label{sec:chapter16_summary}

GRMHD simulations in binary neutron star mergers have rapidly progressed in the last sixteen years. At the initial phase, it was unclear whether the Kelvin-Helmholtz instability could efficiently amplify the magnetic field at the merger, which was originally reported in the Newtonian Smoothed Particle Hydrodynamics simulation~\cite{Price:2006fi}. Currently, the numerical relativity community has a consensus on this picture: {\it the Kelvin-Helmholtz instability develops strongly magnetized turbulence in a short timescale $<5~{\rm ms}$ at the merger.} 
However, the physical saturation of the Kelvin-Helmholtz instability with $B\gtrsim 10^{15-16}~{\rm G}$ needs to be investigated further. 

Also, the way in which the large-scale magnetic field inside the merger remnant builds up needs to be clarified more. A couple of simulations suggest {\it if the large-scale magnetic field is established before the remnant collapses into the black hole, jet launching could be possible.} 
However, the numerical relativity community keeps asking the question: Is it a relic of the large-scale pre-merger magnetic field? Or does a non-trivial physical process work to generate the large-scale magnetic field from the small-scale magnetic field? 
Ultimately, the community has to answer a question: does a merger simulation starting from a {\it strong and large-scale} pre-merger magnetic field with ``standard" grid resolution lead to a physically equivalent outcome of a simulation starting from a {\it weak and large-scale} pre-merger magnetic field with ``high" grid resolution?

A recent demonstration of the MRI-driven $\alpha\Omega$ dynamo inside the remnant massive neutron star could resolve a piece of the puzzle on this issue. 
The lesson from it is {\it that we need an in-depth resolution study and a novel analysis to disentangle the large-scale field generated by the non-trivial process from the relic large-scale field.} With them, it is possible to quantify how the assumed large-scale pre-merger field affects the simulation outcome. Otherwise, it is impossible to reject the possibility that the outcome is merely a consequence of the unrealistically large-scale pre-merger magnetic field. 
It should be also emphasized that the resolution study is essential to build a physical theoretical model, which should be compared to the observables such as AT 2017gfo and GRB170817A because a higher resolution simulation makes more than one order of magnitude difference in the quantities relevant to the electromagnetic counterpart modeling as demonstrated in Ref.~\cite{Kiuchi:2023obe}. Also, this work brings up a new question: {\it What is a realistic time scale to build up the large-scale magnetic field via $\alpha\Omega$ dynamo? Is there any other possibilities for the large-scale dynamo such as the Taylor-Spruit dynamo?} 
Since Ref.~\cite{Kiuchi:2023obe} pointed out that the growth timescale for the large-scale magnetic field generation approximately agrees with the period of the dynamo cycle, i.e., the butterfly diagram. However, the mechanism to set the large-scale field strength just {\it after the merger} is an open question. A conservative estimate based on the merged poloidal magnetic flux of the pre-merger magnetic field with $10^{12}~{\rm G}$, i.e., highly magnetized end of the binary pulsars, gives the time scale of $O(0.1)~{\rm s}$ for the large-scale magnetic field to build up~\cite{Kiuchi:2023obe}. This implies that if a remnant massive neutron star survives longer than this time scale, it could be a central engine of short gamma-ray bursts. The other possibility is the large-scale dynamo inside a torus formed after a remnant massive neutron star collapses into a black hole~\cite{Gottlieb:2023sja}. Once the large-scale field is established inside the torus, the Blandford-Znajek mechanism could drive a relativistic jet from the black hole~\cite{Blandford:1977ds}. 
However, it is an open question whether or not the large-scale field strong enough to extract the black hole rotational energy efficiently is possible via the disk dynamo. The numerical relativity community will continue exploring these possibilities by more sophisticated simulations. 

The numerical relativity community does not have a consensus on the MRI inside the merger remnant massive neutron star. The gradient sub-grid scale model simulations starting from a realistic pre-merger magnetic field strength suggest that the remnant massive neutron star has a high non-axisymmetric intensity, implying the conventional way to estimate $\lambda_{\rm MRI}$ may not be valid~\cite{Aguilera-Miret:2023qih,Palenzuela:2021gdo}, which many simulations starting from the large-scale pre-merger magnetic field rely on to quantify to what extent the simulations resolve the MRI (see Ref.~\cite{Kiuchi:2017zzg} for example). 
However, it does not necessarily mean whether or not the MRI sets in inside the merger remnant, but it means we need to seek a robust way to quantify the MRI's emergence or non-emergence, particularly the MRI's non-linear phase. One potential proposal is to measure the Shakura-Sunyaev parameter or the Maxwell stress. Furthermore, the diagnostics proposed in Ref.~\cite{Hawley:2011tq} could reasonably estimate how the MRI-driven turbulence is sustained. It should be noted that in the non-linear phase of the MRI, the magnetic field is strongly turbulent. It must have a high non-axisymmetric intensity and no static, large-scale background. 
Also, evaluating the mean poloidal field in the MRI active and inactive region could be another way to quantify the emergence of the MRI as demonstrated in Ref.~\cite{Kiuchi:2023obe}. 

The numerical relativity community has a consensus that grid resolution is essential for GRMHD simulations of binary neutron star mergers. However, the {\it effective} resolution could be determined by combining the quality of the Riemann solver and cell-reconstruction scheme. Most of the existing NR codes employ the HLL(E) or its variant LLF Riemann solver with a higher-order cell reconstruction scheme, such as MP5, except for Ref.~\cite{Kiuchi:2022ubj} which employs the HLLD Riemann solver. 
Quantifying the effective grid resolution with the different Riemann solvers and cell-reconstruction schemes is a task that needs to be pursued in the future. 

In summary, numerical modeling of binary neutron star mergers based on GRMHD simulations will continue playing a pivotal role in the multimessenger era. The numerical relativity community will keep making an effort to develop physical modeling for binary neutron star mergers, whose quality is quantitatively good enough to compare to observational data, not only the gravitational waves but also the electromagnetic signals.

\begin{acknowledgement}
 K.K. thanks to the Computational Relativistic Astrophysics members in AEI for a stimulating discussion. 
 K.K. also thanks Luciano Rezzolla, Bruno Giacomazzo, Carlos Palenzuela, Riccardo Ciolfi, Milton Ruiz, Elias Most, Ricard Aguilera-Miret, Luciano Combi, and Philipp M\"{o}sta for their feedback to the chapter. 
 This work is in part supported by the Grant-in-Aid for Scientific Research (grant Nos. 23H01172) of Japan MEXT/JSPS, and by the HPCI System Research Project (Project ID: hp220174, hp220392, hp230204, hp230084, hp240039). 
 
\end{acknowledgement}







\end{document}